\documentclass[twocolumn]{article}
\usepackage[margin=0.55in]{geometry}
\usepackage[T1]{fontenc}
\usepackage[utf8]{inputenc}
\usepackage{lmodern}
\usepackage{graphicx}
\usepackage[percent]{overpic}
\usepackage{multirow}
\usepackage[english]{babel}
\usepackage{subcaption}
\usepackage{caption}
\usepackage{amsmath}
\usepackage{makecell}

\usepackage{numprint}
\npthousandsep{,}

\usepackage{hyperref}
\hypersetup{
    colorlinks=true,
    linkcolor=black,
    citecolor=black,
    filecolor=black,
    urlcolor=black,
    linkbordercolor = {1 1 1},
}

\newcommand{\kph}{\,\mathrm{km}/\mathrm{h}}

\renewcommand{\vec}[1]			% "vectors" are displayed as bold
             {{\mathbf{#1}}}

\usepackage{multicol}

\usepackage{algpseudocode}
\usepackage{algorithm}

\algblockdefx{Step}{EndStep}[1]{#1}{}
\algtext*{EndStep}

\usepackage{CJKutf8}

\usepackage[square,numbers]{natbib}
\title{Estimating the potential for shared autonomous scooters%
\footnote{\copyright 2021 IEEE. Personal use of this material is permitted. Permission from IEEE must be obtained for all other uses, in any current or future media, including reprinting/republishing this material for advertising or promotional purposes, creating new collective works, for resale or redistribution to servers or lists, or reuse of any copyrighted component of this work in other works. DOI:\href{https://doi.org/10.1109/TITS.2020.3047141}{10.1109/TITS.2020.3047141}}}
\author{\normalsize
	D\'aniel Kondor$^{1}$, Xiaohu Zhang (张啸虎)$^{1,2,\ast}$, Malika Meghjani$^{1,3}$, Paolo Santi$^{2,4}$,
	Jinhua Zhao$^{5}$, Carlo Ratti$^{2}$ \\[1ex]
		\normalsize{$^1$ Singapore-MIT Alliance for Research and Technology, Singapore}\\
		\normalsize{$^2$ Senseable City Laboratory, MIT, Cambridge MA 02139 USA}\\
		\normalsize{$^3$ Singapore University of Technology and Design, 487372 Singapore}\\
		\normalsize{$^4$ Istituto di Informatica e Telematica del CNR, Pisa, Italy}\\
		\normalsize{$^5$ Department of Urban Studies and Planning, MIT, Cambridge MA 02139 USA}\\
		\normalsize{$^\ast$ E-mail: \texttt{zhangxh@mit.edu}}}
\date{}

\begin{document}

\begin{CJK*}{UTF8}{gbsn}

\maketitle

\end{CJK*}

\section*{Abstract}
Recent technological developments have shown significant potential for transforming urban mobility. Considering first- and last-mile travel and short trips, the rapid adoption of dockless bike-share systems showed the possibility of disruptive change, while simultaneously presenting new challenges, such as fleet management or the use of public spaces. In this paper, we evaluate the operational characteristics of a new class of shared vehicles that are being actively developed in the industry: scooters with self-repositioning capabilities. We do this by adapting the methodology of shareability networks to a large-scale dataset of dockless bike-share usage, giving us estimates of ideal fleet size under varying assumptions of fleet operations. We show that the availability of self-repositioning capabilities can help achieve up to 10 times higher utilization of vehicles than possible in current bike-share systems. We show that actual benefits will highly depend on the availability of dedicated infrastructure, a key issue for scooter and bicycle use. Based on our results, we envision that technological advances can present an opportunity to rethink urban infrastructures and how transportation can be effectively organized in cities.

\section{Introduction}

	The transportation landscape in cities is changing rapidly, with three important areas of technological advancement driving disruptive changes~\cite{Fulton2017}: (1) connected devices enabling real-time optimizations and new, on-demand transportation modes~\cite{Firnkorn2012,Santi2014,Jorge2013,Lowalekar2017,Shen2018,Xu2019}; (2) improvements in electric propulsion and battery technology resulting in cleaner vehicles and new form factors~\cite{lamprecht2019decentralized,lamprecht2018improving}; and (3) autonomous vehicle technology that promises profound changes with urban mobility~\cite{Smith2012,Harper2016,Kondor2018b,Meghjani2018}. Despite the possibilities afforded by new technologies, there is a lot of uncertainty about how the future of urban transportation will look like, with serious concerns whether sustainable modes of transportation can remain competitive in the age of cheap on-demand autonomous mobility~\cite{Smith2012,Harper2016,Wen2018,Shen2018b,Basu2018,Mo2018}.
		
	A central issue in transportation that has been elusive in the past over hundred years is providing first- and last-mile travel so that commuters can reach high-capacity modes in a convenient and efficient manner. Despite research suggesting that ride-hailing or shared autonomous vehicles (SAVs) could serve this role~\cite{Yan2018,Wen2018,Shen2018b}, there are concerns whether it could work in a scalable and affordable manner~\cite{Yan2018,Jin2018}. Consequently, other form factors beside full-size cars should be considered to further reduce costs, congestion and energy use. Recently, shared bicycles and scooters%
	\footnote{In this article, we use the term \emph{scooter} to refer to a personal mobility device which is suitable to travel on pedestrian path. We specifically limit the term to not include small motorcycles that are often referred to scooters in other contexts.} %
	have been deployed in many cities to provide a sustainable transportation mode for short trips~\cite{Faghih-Imani2017,Shen2018,Xu2019,McKenzie2019,Zhu2020}. While popular, these services face serious challenges since imbalances in demand result in vehicles accumulating in some locations while being unavailable in others; to avoid this, operators are required to spend significant cost and effort on rebalancing the fleet, i.e.~employing people to move vehicles to areas with high demand~\cite{Angeloudis2014,Lowalekar2017,Ghosh2017a,Warrington2019}.
	
	In this paper, we consider a new form of transportation, \emph{self-repositioning shared personal mobility devices} (SRSPMD) as a potential way of providing efficient first- and last-mile transportation and serving short trips~\cite{Garcia2018}. An SRSPMD service would use small electric vehicles, e.g.~scooters, that can move autonomously at slow speed to reposition themselves, but require to be driven by their user during trips. This would allow efficient fleet operations, while keeping the vehicles lightweight and simple.
	
	We perform an evaluation of the benefits of SRSPMDs under the rigorous theoretical framework of vehicle shareability networks~\cite{Vazifeh2018} using real-world data of shared bicycle usage~\cite{Shen2018} and public bus use for short trips in Singapore as our basis. This results in a characterization of ideal SRSPMD fleet size and vehicle utilization required to serve trips currently taken by shared bikes. We compare these results to a simulated scenario of simple fleet management without proactive rebalancing and with limited knowledge of future trips. This way, we provide reasonable bounds of service efficiency for future operators under real-world conditions and evaluate the benefits of predictive fleet management. These results allow us to characterize the main benefits and challenges for SRSPMDs in cities. 
	
	While a significant amount of work has focused on bikesharing, e.g.~on understanding usage patterns~\cite{Shen2018,Xu2019,McKenzie2019}, optimizing fleet rebalancing~\cite{Ghosh2017a,Lowalekar2017,Warrington2019} or determining optimal fleet size~\cite{Lin2011,Celebi2018,Meghjani2018}, it is yet unclear how these results would apply to an SRSPMD operator, where there is no need to perform relocations in batches. Our approach is more similar to previous work focusing on taxi or autonomous vehicle fleet operations~\cite{Vazifeh2018,Kondor2018b}; a main difference is the slow speed of SRSPMD vehicles that can be a serious limitation when estimating which trips can be served consecutively by the same vehicle. To account for this, we explicitly focus on related factors such as (1) state of infrastructure and effect of upgrades; (2) effect of predictions in demand on fleet management; (3) variation of fleet utilization with demand.
	
	Summarizing, the main contributions of this paper are:
	
	\begin{enumerate}
		\item Investigation into the fleet size requirements and vehicle utilization of an operator of shared scooters with self-repositioning capabilities based on real-world data about the demand for short trips.
		\item Explicit characterization of the benefits of limited autonomy, with the average vehicle speed being a main parameter.
		\item Explicit characterization of the benefit of having a knowledge of trips in advance for a limited prediction window.
		\item Explicit characterization of how fleet size and vehicle utilization scales with demand.
		\item Consideration of the benefits of selective infrastructure upgrades.
	\end{enumerate}

\section{The SRSPMD concept}

	There has been significant research in vehicle technology, including lightweight electric propulsion and autonomy. There are prototype autonomous vehicles of small form-factor, including buggies~\cite{pendleton2015autonomous}, wheelchairs~\cite{pineau2007smartwheeler}, and heavy-weight personal mobility devices (PMDs)~\cite{Andersen2016}. These all provide functionality for autonomous driving with a user on board, requiring the vehicles to be sufficiently bulky and heavy so that balancing is possible without active participation of the rider on board -- this is in contrast to truly light-weight vehicles such as bicycles and (kick-)scooters that are balanced by the rider. Beside the issue of balancing, navigating in mixed pedestrian environment presents issues regarding detection and communication of intentions or avoidance of collisions~\cite{Yang2017,Aziz2018,Luo2019}. This can result in autonomous vehicles adopting a ``defensive'' driving style, characterized by frequent braking and overall low speed that makes them unattractive to passengers.
	
	Some of the above issues could be solved by ``hybrid'' vehicles: in this case, the vehicle is driven by its user, but can also move autonomously when a human is not on board. This presents two main advantages: (1) vehicles can be smaller and lighter; (2) autonomous operations can target slower speeds and more defensive driving styles during relocation trips, while during trips with a human, it is up to the rider to manage navigation among pedestrians and other human-driven vehicles. Such vehicles are also likely to be cheaper to develop and manufacture. There has been a significant amount of research to develop electric bicycles with such self-repositioning capabilities~\cite{Wang2012,He2015,Stasinopoulos2017} for the potential use in shared fleets~\cite{Garcia2018}. Similarly, development of such vehicles in a scooter form-factor is actively pursued commercially~\cite{scootbee,ninebot}.
	
	\begin{figure}
		\centering
		\includegraphics[width=1.8in]{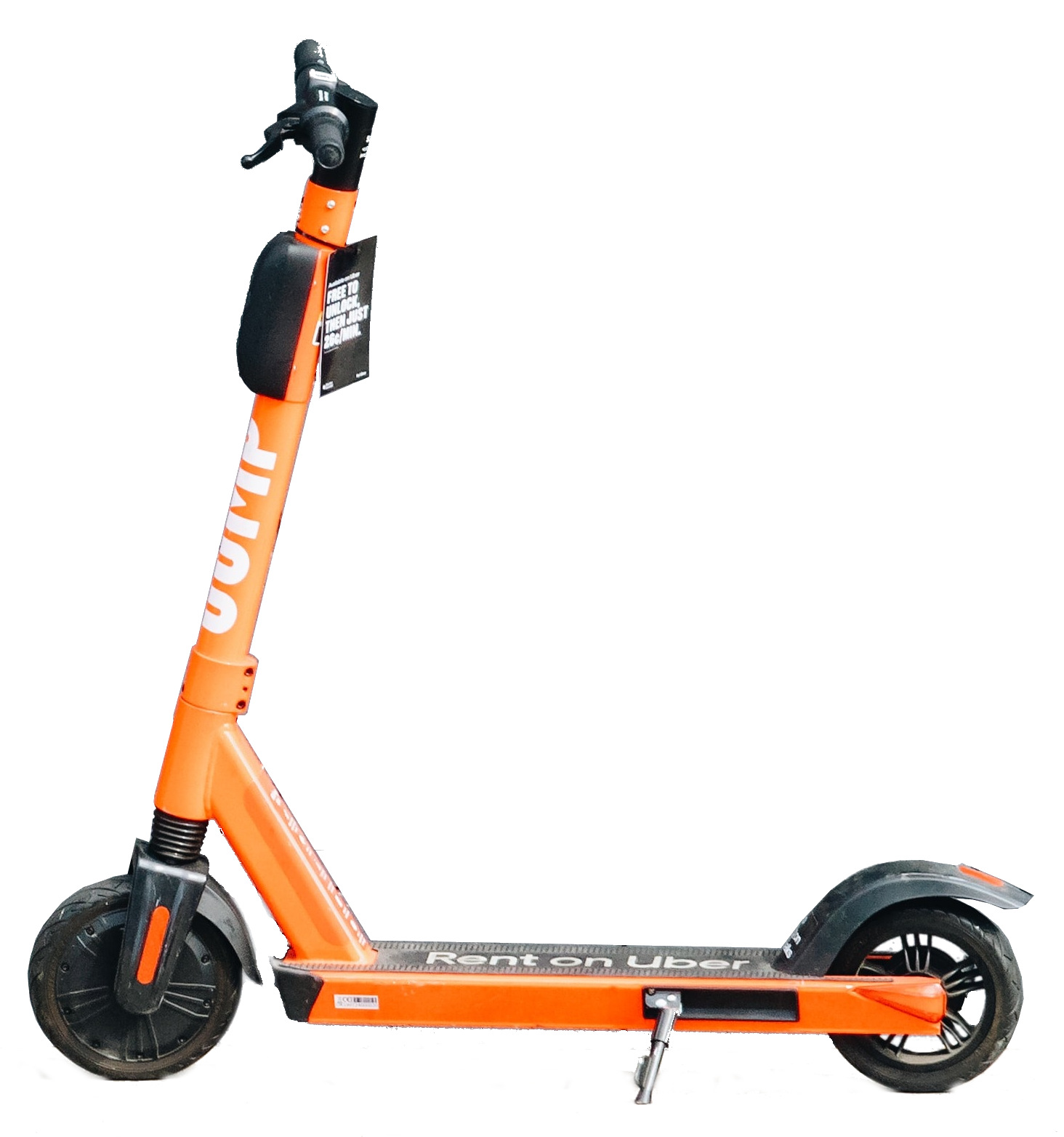}
		\caption{Typical light-weight vehicle that is currently used for shared micromobility services and that could be the basis of an SRSPMD service following hardware upgrades including: (1) at least three wheels for balance; (2) cameras and optionally LIDAR sensors for localization; (3) low-power on-board computer for semi-autonomous navigation; (4) actuators for computer-controlled steering control; (5) optionally, connectors or wireless energy transfer equipment for autonomous charging.}
		\label{scooter1}
	\end{figure}
	
	In the current work, we use the generic term \emph{self-repositioning shared personal mobility device} (SRSPMD) to refer to any such hybrid vehicle, regardless whether it is based on a bicycle or a scooter, such as the vehicle depicted in Fig.~\ref{scooter1}. We note that we consider these vehicles to be strictly separated from full-size cars, i.e.~they are not expected to travel on roads. Also, we only consider electric vehicles, powered by an on-board battery; current shared scooters typically have a range up to 50 km~\cite{Anderson-Hall2019}. We note that our work is naturally extensible to vehicles with full autonomy or to mixed fleets, where the fully autonomous vehicles are provided for users who are unable or unwilling to manually drive them.
	
	We expect SRSPMDs to have multiple advantages compared to conventional shared bicycles or scooters. First, due to autonomous repositioning, they will be able to serve a larger number of trips without the need for costly manual rebalancing; investigating this benefit is the main focus of the current work. Second, having a smaller fleet serving more trips per vehicle, combined with autonomous rebalancing, will avoid the accumulation of idle vehicles in popular destinations, a commonly cited issue with current operators. Lastly, self-repositioning will allow operators to provide a more \emph{reliable} service. Currently, use of shared bicycles or scooters is mainly opportunistic, since there is a large uncertainty about the availability of vehicles. This way, bicycles and scooters remain a secondary, optional choice for their users. With self-repositioning, operators can offer stronger guarantees on the availability of vehicles in the right location and thus users can rely on SRSPMDs as their \emph{primary} mode of transportation for short trips and first- and last-mile trips. In summary, SRSPMD operations will be fundamentally different from current shared vehicles -- either bicycles or scooters -- and could present new solutions to serve short trips in urban environments.

\section{Methods}
	
\subsection{Shared bike usage data}
	
	As the main data source for the current work, we use trips made by customers of a dockless bike-share operator over the one week period between 2017.09.11 and 2017.09.17.~\cite{Shen2018}, available as Ref.~\cite{bikedata}. Data collection and preprocessing procedures were presented in more detail in Refs.~\cite{Shen2018,Xu2019} and in the Supplementary Material. We have a total of 278,826 trips left made by 32,782 unique bikes (identified by the 9 digit unique ID for each bike reported in the dataset). We note that overall, $35.9\%$ of trips either start or end within a $200\,\mathrm{m}$ distance of a rapid transit (MRT or LRT) station exit, indicating that a significant share of trips are first- and last-mile transportation. This observation is consistent with previous work investigating shared bike usage in Singapore~\cite{Shen2018,Xu2019}.
	
	The number of trips and bikes in the dataset indicates that each bike makes on average $1.2151$ trips per day. In reality however, the number of bikes used each day is much lower, between 13,000 and 18,000, thus the average number of trips per bike per day is between $2.3$ and $2.75$ (see Fig.~\ref{fleet_res} and SI Table~\ref{usage_trips_tab}) and on average, each bike is used for $26.2\,$minutes each day (see SI Fig.~\ref{trip_times}). We speculate that the large discrepancy between the total fleet size and daily active fleet is due to multiple factors, including intentional oversupply of bikes in a highly competitive market at the time of our data collection, and bikes being broken or left in hard-to-find locations by users for extended periods of time.

\subsection{Random trips based on bus usage data}

\begin{algorithm}[t]
	\begin{algorithmic}[1]
		\State T: $\{ t_i = (s_{t_i}, e_{t_i}, h_{t_i}), i = 1 \ldots N_T \}$, all trips
		\State B: $\{ b_i = (n_i, s_i, d_i), i = 1 \ldots N_B \}$, all buildings
		\State $n_t$: number of trips to generate
		\State $d_\textrm{max}$: maximum distance of generated trips
		\State $v$: travel speed of trips
		\State R: $\{ \}$ result set of generated trips, initially empty
		\State $i = 0$
		\While{$i < n_t$}
			\State $t =$ Random(T)
			\State B$_S$ = $\{ b_i \in \textrm{B}: s_{b_i} = s_t \}$
			\State B$_E$ = $\{ b_i \in \textrm{B}: s_{b_i} = e_t \}$
			\State $b_s =$ Random(B$_S$)
			\State $b_e =$ Random(B$_E$)
			\State $d = \textrm{dist}(n_{b_s}, n_{b_e}) + d_{b_s} + d_{b_e}$
			\If{$d \leq d_\textrm{max}$}
				\State $T_s = h_t + $ Random($\{ 0, \ldots 3599 \, \mathrm{s} \}$)
				\State $T_e = T_s + d / v$
				\State Insert(R, $(n_{b_s}, d_{b_s}, n_{b_e}, d_{b_e}, T_s, T_e)$)
				\State $i = i + 1$
			\EndIf
		\EndWhile
	\end{algorithmic}
	\caption{Method used to generate candidate SRSPMD trips based on a set of bus trips. The input is the set of bus trips, T, and the set of building, B, downloaded from OSM. The output is the result set of SRSPMD trips, R. The function Random() selects one element from a set in a uniformly random way; the function Insert() adds an element to a set.}
	\label{alg:bustrips}
\end{algorithm}

	As a further data source, we downloaded bus usage data from the Singapore Land Transport Authority's DataMall interface for January 2019~\cite{datamall}. The data includes the monthly total number of trips taken between any two bus stop pairs in Singapore, separated between weekdays and weekends and with a time resolution of one hour. We identify bus stops inside a limited study area based on the Toa Payoh neighborhood (mostly dense residential); we have a total of 94 bus stops (see Fig.~\ref{toapayoh} in the Supplementary Material). We find that the average number of trips on weekdays in the study area is 68,499. Among these, we find that about 28\% could be replaced by a PMD trip of less than 1 km, while 84\% could be replaced by a PMD trip that is less than 2 km long; also, 73.6\% of the bus trips in this area either start or end at the vicinity of a rapid transit station (see Fig.~\ref{bus_mrt_ratio} in the Supplementary Material). For comparison, in the original bike dataset, we have on average 710 trips within the Toa Payoh area per day. This shows that there is a large amount of additional short trips happening that could be served by SRSPMDs beside those that are already made by shared bicycles. Of course, not every bus passenger could switch to using PMDs, but even a small share of bus passengers switching will result in significant demand. Considering bus trips for the whole of Singapore, we find that 13\% of them, or approximately 512 thousand trips per day could be replaced by a PMD trip shorter than 1 km; 48\%, or about 1.9 million trips per day could be replaced by a PMD trip under 2~km.
	
	We created potential ``trips'' for SRSPMD users based on the bus trips, using the methodology shown as Algorithm~\ref{alg:bustrips}. Each bus trip in the original dataset is identified by the start and end bus stops ($s_t$ and $e_t$), and the hour it takes place ($h_t$). We assign random start and end locations to bus trips from the set of buildings in the area, obtained from OpenStreetMap (OSM)~\cite{haklay2008openstreetmap}, represented by the set B, where each building is associated by the OSM node ($n_i$, with distance $d_i$) and the bus stop ($s_i$) that is closest to it. Each iteration of the main loop (lines 9--19 in Algorithm~\ref{alg:bustrips}) selects a random bus trip and selects a building from the candidate sets (B$_S$ and B$_E$) which are closest to the start and end bus stop as the assumed origin and destination of the trip. The total distance of the trips is then calculated and it is only included in the result if this is smaller than the maximum desired distance ($d_\textrm{max} = 2\,\mathrm{km}$ in practice). Trip start time is generated at uniformly random within the start hour of the bus trip and travel time is calculated based on an assumed travel speed of $v = 5\,\mathrm{km}/\mathrm{h}$. The main loop is repeated until the desired number of trips ($n_t$) have been generated. We varied $n_t$ between 100 and 40,000, and for each value, we repeated the trip generation process 100 times. Reported results in the paper (e.g.~those in Fig.~\ref{bustrips}) show the average and standard deviation among these 100 random realizations.
	
	We note that most bus stops occur in pairs, i.e.~on opposite sides of a road, used by opposite directions of the same bus service. For the purpose of generating trips, we merged such pairs of bus stops in a preprocessing step to obtain a result that is not dependent on the direction of bus trips taken by the passengers. We share data used for trip generation along with source code implementing Algorithm~\ref{alg:bustrips} online as Ref.~\cite{bikedata}.

\subsection{Path network based on OpenStreetMap data}
	
	Our methods for estimating fleet size rely on estimating when a vehicle can reach a trip request. We do this by extracting the network of sidewalks and cycle paths from OpenStreetMap~\cite{haklay2008openstreetmap} and using this as the path networks SRSPMDs can navigate on. Since we do not have estimates of vehicle travel speed in real-world conditions, we introduce the parameter $v_R$, the \emph{average} speed that SRSPMDs are able to travel during relocation. We emphasize that in our analysis, $v_R$ is not the actual \emph{travel} speed of the vehicles, but the average speed, i.e.~the total distance of the relocation trip divided by the total time taken; this includes any time spent stopping or slowing down due to traffic interactions, a main limitation while navigating in complex environments~\cite{Yang2017,Luo2019}. This way, we are able to incorporate different assumptions on the infrastructure available to SRSPMDs by varying this parameter. We use low values of $v_R = 1\,\mathrm{km} / \mathrm{h}$ and $2.5\,\mathrm{km} / \mathrm{h}$ as representative of a case where SRSPMDs will continue to use sidewalks, thus are required to carefully navigate among pedestrians, limiting both maximum and average speed for the sake of safety. We further perform our analysis with higher $v_R$ values of $5\,\mathrm{km} / \mathrm{h}$ and $10\,\mathrm{km} / \mathrm{h}$ that represent scenarios where SRSPMDs can perform an increasing share of their relocation trips on a path infrastructure separated from pedestrians~\cite{Bao2017,Aziz2018}.

	When using the path network, we map each trip start and end location to the closest OSM node (denoted by $n_{st}$ and $n_{et}$ in the following), noting the Euclidean distance as well ($d_{st}$ and $d_{et}$). When calculating distances between trip start and end locations, we add these extra distances to the shortest path distance calculated along the path network for a more realistic estimate of reachability.
	
	Furthermore, we consider a ``two-tiered'' infrastructure where parts of the OSM path network are upgraded to be better suitable for SRSPMDs and thus allow an increased relocation speed of $v_R^* = 15\,\kph$. We formally define such upgrades by introducing a binary vector $\vec{u}$, where $u_i = 1$ if the $i$th edge is upgraded and $0$ otherwise. We can then define the scalar function $C(\vec{u})$ and $B(\vec{u})$ as the cost and benefit of a specific upgrade configuration. A good upgrade configuration is such that will maximize the benefit for a given cost. In the general case, finding a solution for this problem is compuationally challenging; in this paper, we use an approximation where (1) cost is defined proportional to the total lenght of upgraded paths; (2) benefit is defined as the share of total travel distance on upgraded paths; and (3) instead of finding an exact solution to the maximization problem, we select the top $N$ most frequently used edges as candidates for the upgrade, for $N$ chosen suitably such that the total relative length of upgraded path is equal to a desired ratio. We present a formal definition and more detailed discussion of the path upgrade problem in the Supplementary Material, including the motivation of our choice of upgrades.

\subsection{Oracle model for estimating minimum fleet size}
	
\begin{algorithm}[t]
	\begin{algorithmic}[1]
		\State N: OSM path network
		\State T: $\{ t : (n_{st}, d_{st}, T_{st}, n_{et}, d_{et}, T_{et}) \}$ all trips
		\State $v_R$: vehicle relocation speed
		\State S: $\{ n_s \rightarrow S_n : \{ (n_s, d_i, T_{si}) \}, i = 1, \ldots N_{S_n} \}$
		\State E: $\{ n_e \rightarrow E_n : \{ (n_e, d_i, T_{ei}) \}, i = 1, \ldots N_{E_n} \}$
		\State P: $\{ \}$ set of candidate pairs
		\ForAll{$n_e \in $ E}
			\ForAll{$(d, n) \in $ Dijkstra(N, $n_e$)}
				\ForAll{$s \in $ S($n$), $e \in $ E($n_e$)}
					\State $d_\textrm{tot} = d + d_s + d_e$
					\State $T = d_\textrm{tot} / v_R$
					\If{$T_e + T < T_s$}
						\State Insert(P, $(e, s, d_\textrm{tot})$)
					\EndIf
				\EndFor
			\EndFor
		\EndFor
		\State R = MaximumWeightedMatch(P)
		\ForAll{$(e, s) \in $ R}
			\State $t_e$ = Remove(T, $e$)
			\State $t_s$ = Remove(T, $s$)
			\State Insert(T, Chain($t_e$, $t_s$))
		\EndFor
		\State Result: $|$T$|$, the number of total vehicles used
	\end{algorithmic}
	\caption{Algorithm used to calculate an ideal dispatching in the oracle model for fleet size estimation. The input is the path network, N, compiled from OpenStreetMap data, the set T, that initially contains all trips, and $v_R$, the speed of the vehicles in self-repositioning mode.}
	\label{alg:oracle}
\end{algorithm}

	We use the methodology of shareability networks~\cite{Vazifeh2018} to estimate a theoretical minimum for the fleet size. We use the list of trips as the input, and require all trips to be served by the fleet of SRSPMDs without any delay. We outline the necessary steps for this as Algorithm~\ref{alg:oracle}.
	
	For better performance, we map trip start and end events to nodes in the path network (S and E in Algorithm~\ref{alg:oracle}). For each node that has trip end events, we perform a standard Dijkstra-search (represented by the Dijkstra() function). For each node in the result set ($n$), we evaluate all pairings with the end events at the search start node and start events at $n$ in the loop in lines 9 -- 15. Feasible pairs of events are added to the candidate set P, forming a weighted bipartite network. We calculate a maximum weighted matching (line~18)~\cite{HopcroftKarp, lemon} and use pairs in the results set R to create an ideal dispatching strategy~\cite{Boesch1977, Vazifeh2018}. Individual trips that are present among these pairs are identified and removed from the set of trips T; chains are created and inserted in their place (lines 19--22; functions Remove(), Chain() and Insert() represent these steps). After this procedure, the number of elements (trips and trip chain) in T corresponds to the number of vehicles needed.
	
	We display an overview of runtimes and the size of the shareability networks in Tables~\ref{matching_runtimes} and~\ref{matching_runtimes_limit_tab} and in Fig.~\ref{matching_runtimes_limit_fig} in the Supplementary Material. Since this methodology require advance knowledge of all trips, we call this an \emph{oracle} model, using the terminology of Santi.~et~al~\cite{Santi2014} and Vazifeh~et~al.~\cite{Vazifeh2018}. Major differences from the implementation in Ref.~\cite{Vazifeh2018} include: (1) instead of pre-calculating travel times among a set of nodes, we integrate finding candidate trip pairs with a Dijkstra-search of the path network. This is feasible as we don't expect large variations in travel times during the day, while on the other hand, in accordance with the nature of micromobility, we use a finer-grained spatial resolution of network nodes, i.e.~a total of 79,771 nodes are included in our path network. Storing a travel time matrix among all of them is impractical, while our integrated search approach offers adequate performance. (2) Trip connections in the current work are not limited to short time intervals (the authors of Ref.~\cite{Vazifeh2018} used a $\delta = 15\,\mathrm{min}$ maximum connection time to limit the problem size and deadheading). By allowing longer connections, we are able to exploit more sharing opportunities even when vehicle relocation speeds ($v_R$) are limited; this is essential for micromobility, since limiting the connection time would severly limit the feasible solutions. At the same time, we limit deadheading by incorporating relocation travel as weights in the maximum matching problem. 
	
	The result of this estimation then shows the potential for efficient fleet management under ideal conditions, i.e.~it gives the smallest possible number of vehicles to serve the given number of trips without delay. By comparing the case with and without self-repositioning, we can characterize the maximum potential benefit of autonomy. We note however, that even in the case of an ``oracle'' model, this solution is ideal only if we constrain ourselves by taking the trip start times as fixed; further optimizations are possible if trip start times can vary in an interval, leading to the well-known dial-a-ride problem~\cite{Berbeglia2010}, a central problem in operations research where obtaining exact optimal solutions is computationally infeasible for realistic problem sizes~\cite{Bertsimas2019}.

\subsection{Online model for estimating operational characteristics}
	
\begin{algorithm}[h!]
	\begin{algorithmic}[1]
		\State N: OSM path network
		\State T: $\{ t : (n_{st}, d_{st}, T_{st}, n_{et}, d_{et}, T_{et}, T_{Dt} = 0) \}$ all trips
		\State $v_R$: vehicle relocation speed
		\State $v_W$: walking speed
		\State $d_W$: maximum walking distance
		\State $t_B$: size of batches
		\State $t_W$: maximum wait time
		\State $t_W^* = \min (t_W, d_W / v_W)$ maximum walking time
		\State $d_W^* = d_W + v_R t_W^*$ combined walking distance
		\State $t = 0$ current time
		\State V: $\{ \}$ set of available vehicles (initially empty)
		\While{$|$T$|$ > 0}
			\State E $ = \{ t \in $T$, T_{et} \in [t, t + t_B)$ \} %]$
			\State S $ = \{ t \in $T$, T_{st} \in [t, t + t_B)$ \} %]$
			\ForAll{$t \in $ E}
				\Comment Process trip end events
				\State Insert(V, $(n_{et}, t_v = T_{et} + d_{et} / v_R + T_{Dt})$)
				\State Remove(T, $t$)
			\EndFor
			% \State $P_T = $ FindPairs(V, $T_s$, $t_W$)
			\State $P_T = \{ \}$ \Comment match candidates
			\ForAll{$t \in $ S} \Comment process trip start events
				\ForAll{$v \in $ V}
					\State $d_1 = d(n_v, n_{st}) + d_{st}$
					\State $t_1 = \min(d_1, d_W^*) / (v_R + v_W)$
					\State $t_2 = \max(d_1 - d_W^*, 0) / v_R$
					\State $t_s^* = \max(t_v, t) + t_1 + t_2$
					\If{$t_s^* \leq T_{st} + t_W$}
						\State Insert($P_T$, $(v, t, t_1 - T_{st})$)
					\EndIf
				\EndFor
			\EndFor
			
			\State $R_T = $ MaximumWeightedMatch($P_T$)
			\ForAll{$(v, t, d) \in R_T$}
				\State Remove(V, $v$)
				\State Remove(S, $t$)
				\State $T_{Dt} = d$
			\EndFor
			\ForAll{$t \in $ S}
				\State A new vehicle is added at $n_{st}$ to serve this trip
			\EndFor
			\State $t = t + t_B$
		\EndWhile
		\State Result: $|$V$|$, the number of total vehicles used
	\end{algorithmic}
	\caption{Algorithm used to evaluate online fleet operations and estimate a fleet size needed to serve trips without an advance knowledge of demand. The input is the set of trips to be served (T), the path network (N) and the parameters for vehicle relocation speed ($v_R$), batch size ($t_B$) and maximum wait time ($t_W$).}
	\label{alg:online}
\end{algorithm}

	In previous work applied to taxi data, Vazifeh~et~al.~\cite{Vazifeh2018} showed that an online version of maximum matching can offer similar performance to the oracle model with only short delays. We note that a fundamental difference in the case of SRSPMDs is the slow relocation speed of vehicles that can hamper performance. It is thus important to investigate an online model, where the operator does not have advance knowledge of trip requests.
	
	We use a combination of greedy heuristics and batched maximum matching in short time windows~\cite{Kondor2018b,Vazifeh2018} to simulate the performance of a fleet operator with a simple operating strategy that only includes response to user requests. The main methodology is outlined as Algorithm~\ref{alg:online}; variations that enable longer waiting or look-ahead times and passengers walking are discussed in the Supplementary Material. We share the source code for this model online as Ref.~\cite{online_model_code}.
	
	In this case, trip requests (trip start events) are aggregated in $t_{B} = 1\,\mathrm{min}$ time windows along with trip end events (lines~13 and~14 in Algorithm~\ref{alg:online}). For each time window, we perform a maximum weighted matching between trip start events and available vehicles, that are kept track of in the set V. The generation of the candidate set for this is shown as a double loop in lines~20--30 for a simpler presentation; in practice, we obtain better performance using a limited index or a Dijkstra-search similarly to Algorithm~\ref{alg:oracle}. In the main implementation, the candidate set $P_T$ only includes matches that allow trips to be served with a maximum delay of $t_W = 5\,\mathrm{min}$; for trips that would go unserved, we add new vehicles, increasing the fleet size~\cite{Kondor2018b,Bauer2018}. Maximum matching is performed in a way to minimize the total waiting time of served trips. Actual delay values are saved along the trips (in set T, in line~35) and used when considering the availability of vehicles at the end of trips (lines~15--18) and when matching vehicles to new trips (line~25). The result of this estimation is the number of vehicles in the V set, required to serve all trips.
	
	\begin{figure*}[t]
		\centering
		\begin{minipage}{3.2in}
			\begin{overpic}{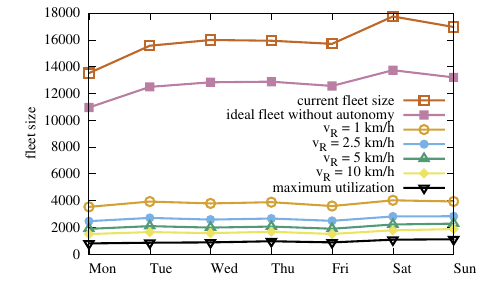} \put(3,64){\large \textsf{A} } \end{overpic} \\
			\hspace*{0.16in}\begin{overpic}{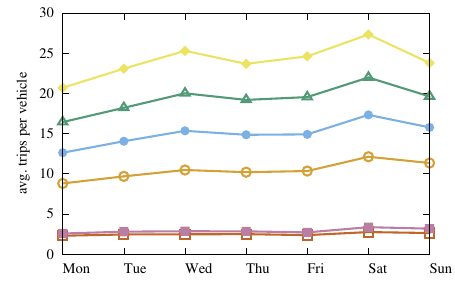} \put(3,64){\large \textsf{C} } \end{overpic}
		\end{minipage}
		\begin{minipage}{3.2in}
			\begin{overpic}{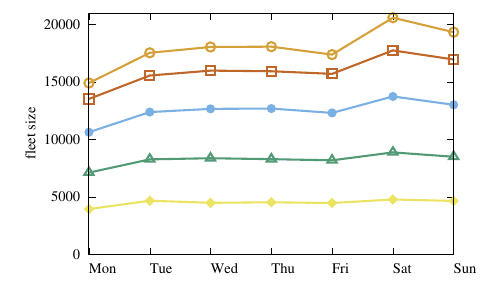} \put(3,64){\large \textsf{B} } \end{overpic} \\
			\hspace*{0.16in}\begin{overpic}{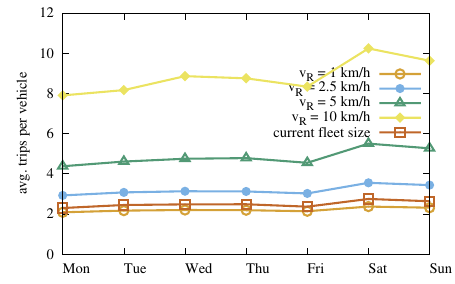} \put(3,64){\large \textsf{D} } \end{overpic}
		\end{minipage}
		
		\caption{Main results for fleet size, vehicle utilization and passenger waiting times. Top row: fleet sizes for the oracle modes (A) and online mode (B) are compared over the course of seven days for different average relocation speed of SRSPMDs ($v_R$). We further show the current fleet size (green line; only bikes that are used at least once that day are counted), the result of an optimal allocation of bikes without autonomy and a walking distance of up to $100\,\mathrm{m}$ for passengers (dark blue line) and the maximum number bikes in use at any time during the day (black line). Note that fleet sizes in the online model can be larger than the current bikeshare fleet size as a result of uncertainties in GPS data and the filtering procedure we applied to the raw data as described in the Materials and Methods section. On the contrary, when calculating an ideal fleet size without autonomy, we accounted for these by keeping the real sequence of trips for each bike as an additional possibility, even if that required larger connection distances. Bottom row: average fleet utilization (i.e.~number of trips per vehicle per day) in the oracle mode (C) and online model (D).}
		\label{fleet_res}
		\label{fleet_online_res}
	\end{figure*}
	
	We note that Algorithm~\ref{alg:online} includes the option that passengers are willing to walk a short distance, up to a $d_W$ limit, to ``meet'' a vehicle to start their trip, with a walking speed of $v_W$. This is a departure from earlier works focusing on car-based services that assumed fixed trip start locations~\cite{Kondor2018b, Vazifeh2018}, and is similar to the work of Meghjani~et~al.~\cite{Meghjani2018}. In this case, passengers spend a maximum $t_W^* = d_W / v_W$ time walking; we include this time in the waiting time and require that $t_W^* \leq t_W$. Thus, when calculating the time needed for a vehicle reaching the start of a trip, in lines~22--25, we include this factor in the calculation via the combined distance $d_W^* \equiv d_W + v_R t_W^*$ that is the distance covered by the passenger and the vehicle together in $t_W^*$ time. The main results of this paper however do not include passengers walking, i.e.~correspond to the case when we set $d_W = 0$ (in this case, we define $t_W^* \equiv 0$ as well).
	
	Two further variations of Algorithm~\ref{alg:online} were used and are presented in the Supplementary Material: (1) instead of ``creating'' new vehicles, we investigate the performance of a fixed fleet, distributed randomly at the beginning of the day, without limiting maximum waiting times (Algorithm~\ref{alg:online_nomax}); (2) a ``limited oracle'' model, where the operator is assumed to have knowledge of trips in a look-ahead window $T_{LA} > t_B$ (Algorithm~\ref{alg:online_tw}).
	
	The main motivation of the limited oracle model is to explore how intelligence or predictions about upcoming trips affect performance. The main limitation of the online model is that we are not considering strategic decisions made by the operators to rebalance the fleet of vehicles that can affect the performance drastically~\cite{Lowalekar2017}. We expect that real operating conditions will present ample such opportunities, as commute patterns are highly regular~\cite{mobility,Toole}. Even without actual predictions, an operator can make rebalancing movements with the aim of ensuring an even spatial distribution of vehicles in the service area. This can drastically improve the performance of the system~\cite{Ghosh2017a,Warrington2019}, however, such strategic methods are demand and scenario dependent and thus not addressed in this work. Instead, by considering advance knowledge of trips in a larger time window, and varying the size of this window, allows us to quantify the maximum benefit from such predictive rebalancing and explore cases in-between the oracle and online models.

\section{Results}

\subsection{Oracle model}

	We display main results for lower and upper bounds on fleet size in Figs.~\ref{fleet_res}A-\ref{fleet_res}D. Further details are given in Tables~\ref{usage_trips_tab}--\ref{tab_distances2} in the Supplementary Material. Ideal fleet sizes in the oracle model range from around 4,000 vehicles for $v_R = 1\,\mathrm{km} / \mathrm{h}$, to between 1,500 and 2,000 for $v_R = 10\,\mathrm{km} / \mathrm{h}$. These present 4 to 10 times reductions compared to the number of active bicycles each day of the bikeshare operator which ranges between 13,500 and 18,000 and up to 17 times reduction compared to the total number of bikes seen in the fleet over the course of one week. At the same time, average daily travel per vehicle is still limited to below 40~km (see SI Table~\ref{tab_distances2}), well within the capabilities of commercial scooters, indicating that any extra costs due to charging infrastructure will be limited.
	
	To better estimate the benefits and limits of self-relocation, we perform two comparisons in the oracle model. First, we estimate an ideal fleet size without autonomy. We do this by assuming stationary vehicles and the willingness to walk up to $d_{walk} = 100\,\mathrm{m}$ by users to reach a bicycle. This corresponds to a case where the operator assigns a bicycle to each user for their trip based on the results of an ``oracle'', instead of the user freely choosing any available bike. We see in Fig.~\ref{fleet_res} that this result offers only moderate improvements in fleet size over the base case, thus we can conclude that self-relocation capabilities are essential for making significant improvements in fleet size and vehicle utilization. We also calculate an absolute minimum on fleet size as the maximum number of bicycles in use simultaneously; this results in very low numbers, between 800 and 1,110.
		
	\begin{figure*}
		\centering
		\begin{overpic}{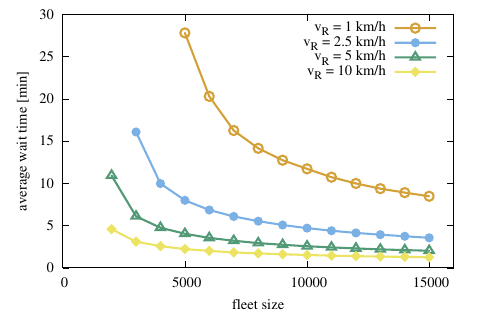}\end{overpic}
		\begin{overpic}{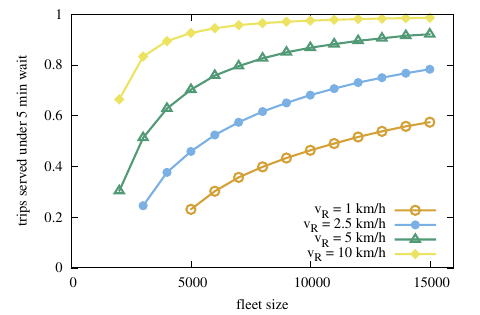}\end{overpic}
		\caption{Average waiting times (left) and ratio of trips served under $t_w = 5\,\mathrm{min}$ waiting time (right) in the online model.}
		\label{fleet_online_res2}
	\end{figure*}

\subsection{Online model}
	
	Having estimated theoretical minimum fleet sizes in the oracle model, we compare these with the upper bounds obtained in the online model. We perform two variations to obtain (1) an estimation of ``ideal'' fleet size without knowledge of trips in advance; (2) a characterization of service quality in terms of waiting time for users. In the first case, we start the simulation with zero vehicles and allow the operators to ``create'' new vehicles when a trip request would go unserved for $t_{w} = 5\,\mathrm{min}$, similarly to the methodology used to estimated SAV fleet sizes previously~\cite{Kondor2018b,Bauer2018}. This results in significantly larger fleet sizes (Fig.~\ref{fleet_online_res}B), comparable to the original fleet size of bicycles for low values of $v_R$ and a more reasonable number of between 4,000 and 5,000 if vehicles can travel faster ($v_R = 10\,\mathrm{km} / \mathrm{h}$). In the second case, we run the simulation with a predetermined number of vehicles distributed randomly in the city and record average waiting times and the ratio of trips served under $t_{w} = 5\,\mathrm{min}$. We can make similar conclusions as in the previous case: in Figs.~\ref{fleet_online_res2}E-\ref{fleet_online_res2}F, we again see that a fleet size between 4,000 and 5,000 vehicles and high $v_R$ values are necessary for adequate service, e.g.~considering a fleet size of 5,000 vehicles, for $v_{R} = 10\,\mathrm{km}/\mathrm{h}$, we have an average waiting time of $2.2\,\mathrm{min}$ and 92.6\% of trips are served within 5 minutes.
	
	In the online model, the operator needs to be ready to serve any trip request with only a small delay; if trip requests are not known in advance, an idle vehicle needs to be available at most $t_w$ travel distance from any location in their service area. For $v_{R} = 1\,\mathrm{km}/\mathrm{h}$ and $t_w = 5\,\mathrm{min}$, this distance is only $83\,$meters; obviously, this translates into a requirement of having a large number of such stand-by vehicles.
	
	By drawing a $100\,\mathrm{m}$ circle around every trip start location in the dataset and merging these, we obtain an estimate of $312\,\mathrm{km}^{2}$ as the service area of the dockless bike share operator in Singapore. For $v_{R} = 1\,\mathrm{km}/\mathrm{h}$, we would need at least $N_I = $22,464 stand-by vehicles to serve any trip request within $t_{w} = 5\,\mathrm{min}$. Obviously, $N_{I} \sim v_{R}^{-2}$; with larger relocation speeds, a smaller number of vehicles is needed to cover the service area: with $v_{R} = 2.5\,\mathrm{km}/\mathrm{h}$ we already only need 3,594 such vehicles, for $v_{R} = 5\,\mathrm{km}/\mathrm{h}$ we need 899 vehicles and for $v_{R} = 10\,\mathrm{km}/\mathrm{h}$ we need 225 vehicles. In reality, available vehicles are not evenly distributed in the service area, nor is the demand. Furthermore, we have to account for the vehicles engaged in serving trips or relocating beside $N_I$. Empirically, we find a lower exponent of about $0.87$ when we consider fleet sizes necessary to serve at least 50\% of trips with a maximum of $5\,\mathrm{min}$ waiting time (see Figure~\ref{online_5min_fleet} in the Supplementary Material).
	
	Results so far were based on the assumption that trips have a fixed start location where passengers are waiting for a vehicle that is used for the trip. In the case of short trips however, it makes sense to assume that users would start their trip walking and continue the trip using an SRSPMD vehicle that they encounter at a suitable ``meeting point''. This resembles how shared vehicles without autonomy operate currently, where users have to find a nearby vehicle. A willingness of passengers to walk a short distance effectively increases the radius where available vehicles can come from and thus lower the $N_I$ idle vehicle number needed to cover an area. We display results for a walking speed of $v_W = 3.6\,\kph$ in Fig.~\ref{bikeshare_online_walk_main} as the change in fleet size and in Fig.~\ref{bikeshare_online_walk} in the Supplementary Material for vehicle utilization. We find that a maximum walking distance of $100\,\mathrm{m}$ results in decreasing the necessary fleet size by over 25\% for $v_R = 1 \kph$. We see that passengers walking has the largest effect for slow vehicle relocation speeds, but is still significant even in the case of higher speeds as it results a 10\% decrease of necessary fleet size even for $v_R = 10 \kph$. Even higher decreases in fleet size are possible for larger maximum walking distances. These results are consistent with the findings of Ref.~\cite{Meghjani2018} who investigated the performance of a fleet of multi-class vehicles with optional walking in a limited geographic area and with synthetic demand data.
	
	\begin{figure}
		\centering
		\includegraphics{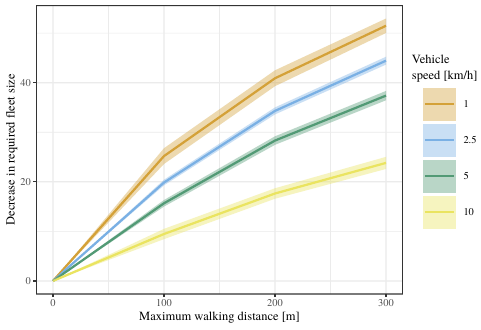}
		\caption{Decrease in fleet sizes achievable if passengers are willing to walk short distances to meet a vehicle. Results are shown for the online model as a function of the maximum walking distance acceptable to passengers. We show results averaged over the seven days of data; ribbons show the standard deviation over these.}
		\label{bikeshare_online_walk_main}
	\end{figure}	
	
	\begin{figure*}
		\centering
		\includegraphics{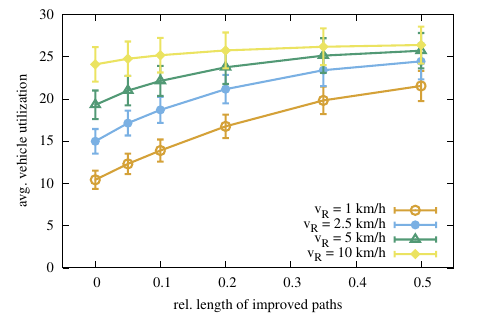}
		\includegraphics{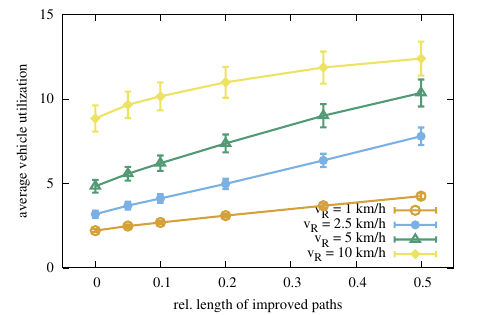}
		\caption{Average utilization of SRSPMD fleet in the oracle model (left) and online model (right) after improving a given relative length of paths. Results are averaged over the 7 days of data; separate results for each day are shown in the Supplementary Material.}
		\label{improvements2}
		\label{online_improvements}
	\end{figure*}
	
	\begin{figure*}[t]
		\centering
		\begin{overpic}{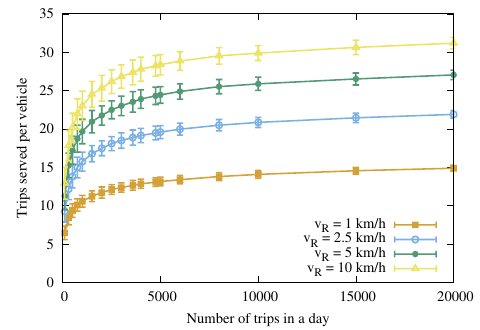} \end{overpic}
		\begin{overpic}{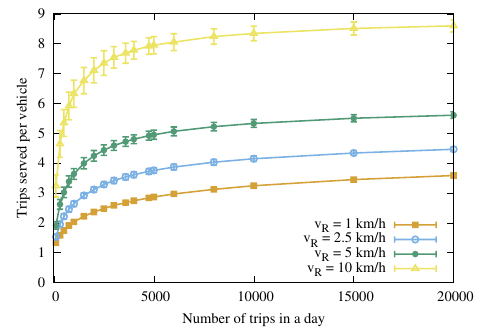} \end{overpic}
		\caption{Average fleet utilization as the function of the number of daily trips in Toa Payoh, with trips generated based on bus usage data. 
		Left: oracle model; right: online model.}
		\label{bustrips}
	\end{figure*}
	
	We anticipate that real operating conditions will present a middle ground between the oracle and online models; operators will likely be able to make valuable predictions on the demand for trips. We carried out an estimation for the maximum gains of such predictions in a ``limited oracle'' model, where trip requests are assumed to be perfectly known in advance in a time window $T_{LA}$ and vehicle assignments are recalculated periodically based on the new information that is available (see Algorithm~\ref{alg:online_tw} in the Supplementary Material). In Figure~\ref{bikeshare_online_tw1} in the Supplementary Material, we show improvements in average vehicle utilization as a result of increasing the $T_{LA}$ look-ahead window size. Vehicle utilization increases gradually with relocation speed and approaches that of the oracle model for $T_{LA}$ above four hours. Notably, for the slowest speed, $v_R = 1 \kph$, utilization increases only slowly, showing that slow repositioning speed will hamper efficient operations even if predictions about the demand are available.
	
\subsection{Effects of upgrading infrastructure}
\label{sec:improvements}

	Our analysis so far outlines that the $v_R$ average relocation speed plays a crucial role in the viability of an SRSPMD service, especially in the online model. To further characterize the benefits from upgrading path infrastructure, we repeat our previous analyses in a presumed ``two-tiered'' infrastructure system: in this case, we have separate paths upgraded specifically for PMD, SRSPMD and potentially bicycle use, allowing high average relocation speed of $v_R^* = 15\,\mathrm{km} / \mathrm{h}$. The ratio of such paths among all is controlled by the parameter $r \in \{0.05, 0.1, 0.2, 0.25, 0.5\}$. On the rest of the path network, we assume the same travel speeds as previously, controlled by the $v_{R}$ parameter. We envision path upgrades starting with the most used segments, continuing by decreasing usage rank until a total of $r$ fraction of path length is reached (we present a more thorough discussion of the upgrade problem in the Supplementary Material).
	
	We display results for average vehicle utilization using the two-tier infrastructure in Fig.~\ref{improvements2}. We see that significant improvements in utilization are possible for relatively minor upgrades in infrastructure. These increases in average vehicle utilization correspond to decrease in total fleet size; more detailed results are  displayed in Figs.~\ref{improved_fleetsize1}--\ref{improved_online_ntrips1} in the Supplementary Material.
	
\subsection{Scaling of utilization with demand}
	
	Our main analysis was carried out using data from a bikeshare operator as an estimate of the demand for short trips, but is in itself limited by the usage patterns and market share of the operator. To overcome this limitation, we performed an additional analysis in which we sampled trips made by bus passengers in Toa Payoh, a dense residential neighborhood of Singapore (see map of study area in Fig.~\ref{toapayoh} in the Supplementary Material). This analysis allows us to answer the question ``If there were twice as many trips to serve, do we need twice as many vehicles, or is there an econmies of scale effect that allows more efficient operations as demand grows?''.
	
	More specifically, we characterize the dependence of average vehicle utilization in an ideal fleet on the total demand by creating random trips based on the bus data and varying the $n_t$ number of daily trips. For each $n_t$ value, we repeat our estimate of fleet size in the oracle and online models and calculate the average number of trips served by a vehicle in the fleet. In Fig.~\ref{bustrips}, we display this measure as a function of the number of daily trips, $n_t$. We see that initially there is a strong dependence between these, showing that serving a higher demand is possible in a more efficient way. Essentially, as the trips become more dense, there are more connection opportunities for the vehicles. However, we see there is a limit: above 5,000 trips per day, vehicle utilization seems to saturate and there are only marginal benefits of more demand. In Fig.~\ref{toapayoh} in the Supplementary Material, we show that below this saturation, average utilization of vehicles can be well modeled to grow logarithmically with the number of daily trips. Comparing to the results for the bikeshare data (in Fig.~\ref{fleet_res}), we see that there is a moderate room for further improvement if higher usage rates are achieved.
	
	Comparing these results with the same calculated using the bikeshare data limited to the Toa Payoh area, we see in Fig.~\ref{bustrips_2} in the Supplementary Material that in the case of the oracle model, the two fit together well, thus we can regard the results from the bus trip data as a meaningful extrapolation. Conversely, we find that there is no such correspondence in the case of the online model; there, results based on bikeshare data show much higher vehicle utilization than results based on bus trips, values that are only reached when a much higher volume of bus trips are considered. One possible explanation for this is that the bikeshare trips are already biased by the availability of vehicles, since we have no data on lost demand. This highlights the difficulties likely encountered by an operator providing an on-demand mobility service.
		
\section{Discussion}

	We note that there is a clear cost component of SRSPMDs. We estimate the average cost of conventional electric scooters approved by the Land Transport Authority of Singapore as 566~SGD~\cite{scooterprice} and the cost of an autonomous version to be about thrice as much, i.e.~around 1,500~SGD (including the cost for a short range LIDAR and an on-board computer); similar price estimates are given by manufacturers actively developing scooter models with limited autonomy~\cite{thedrive}.
	This implies that deploying a fleet of SRSPMDs will be financially reasonable over convential vehicles if the fleet size is reduced to one third or less. Considering that the average active fleet size of the bike sharing operator is about 15,000, this is true for all cases in the oracle model that we investigated and also for the online model if an average relocation speed of $v_R = 5\,\mathrm{km} / \mathrm{h}$ or higher is achieved (see also Fig.~\ref{fig:fleet_size_cost_wt} in the Supplementary Material). Beyond capital costs, further factors will include extra liability insurance and maintenance requirements for the SRSPMDs, while at the same time, significant savings can be realized due to eliminating the manpower needs for fleet rebalancing. Both conventional and autonomous scooters are expected to incur similar costs for battery charging or swapping; at the same time, automated solutions for charging could realize further savings for operators.

	While in the current work, we used the $v_R$ reposition speed as a parameter, in reality it will be determined by the ability of the vehicles to navigate in a complex environment. This way, operations can be severely affected if SRSPMDs have to share narrow sidewalks with pedestrians~\cite{Yang2017,Aziz2018,Luo2019}. Differences between the oracle and online models presented in the current work also highlight the need for \emph{predictive} repositioning instead of \emph{reactive} fleet management, especially in areas where repositioning speed remains slow. At the same time, more infrastructure will be also needed to avoid conflicts among pedestrians, SRSPMDs, and cyclists. Future research should focus on the design for sidewalks, paths and crossings to minimize such conflict and ensure safety for all transportation modes. This will require extending the limited research on cyclist behavior and maneuvering~\cite{Chen2018,Mohammed2019} for the case of human-driven and self-repositioning scooters.
	
	Further research should consider the full sustainability benefits of SRSPMDs, including lifecycyle energy use under different scenarios of usage patterns and integration with public transit services~\cite{Shen2018b}. We believe the main potential for positive change is solving the first- and last-mile transportation problem. SRSPMDs can effectively increase the catchment area of rapid transit stations~\cite{Guerra2012}, and relieve buses and road capacity from short trips. Increased convenience can help transit remain a competitive choice for travel; this is especially important considering a future with autonomous cars offering cheap point-to-point transportation~\cite{Basu2018}. At the same time, SRSPMDs have the potential to replace active modes (i.e.~walking and cycling), highlighting the need for a holistic approach to evaluate the full effect on the sustainability of urban transportation.
	
	Looking beyond, we believe that deployment of SRSPMDs and the implied infrastructure needs should be studied together with the opportunities offered by the three main technological advances in transportation, i.e.~connected devices, electric mobility, and autonomy. The combination of these offers us the opportunity to rethink the design of transportation infrastructure in cities, a change that can be compared to the effect that the internal combustion engine and electric rail transit had on cities more than a hundred years ago. Competition between private cars and mass transit could be transformed into the management of a more fluid landscape of shared, connected, electric and autonomous transportation solutions of various form factors and operational models, accompanied by and evolution of the transportation network infrastructure to support them.

\section*{Acknowledgements}

This research is supported by the Singapore Ministry of National Development and the National Research Foundation, Prime Minister’s Office, under the Singapore-MIT Alliance for Research and Technology (SMART) programme.

We thank Allianz, Amsterdam Institute for Advanced Metropolitan Solutions, Brose, Cisco, Ericsson, Fraunhofer Institute, Liberty Mutual Institute, Kuwait–MIT Center for Natural Resources and the Environment, Shenzhen, UBER, Victoria State Government, Volkswagen Group America, and all of the members of the MIT Senseable City Laboratory Consortium for supporting this research.

\small

\bibliographystyle{unsrtnat}

\newpage
\normalsize

\twocolumn[  
    \begin{@twocolumnfalse}
    
		\begin{center}
			{\huge \bf Supplementary Material \\[2.5ex]}
		\end{center}
		
    \end{@twocolumnfalse}
]

\renewcommand{\thefigure}{S\arabic{figure}}
\renewcommand{\thetable}{S\arabic{table}}
\renewcommand{\thealgorithm}{S\arabic{algorithm}}

\setcounter{figure}{0}
\setcounter{table}{0}
\setcounter{algorithm}{0}

\makeatletter
\let\ftype@table\ftype@figure
\makeatother

\subsection{Preprocessing for the bike trip data}
	
\begin{figure*}
	\centering
	\includegraphics{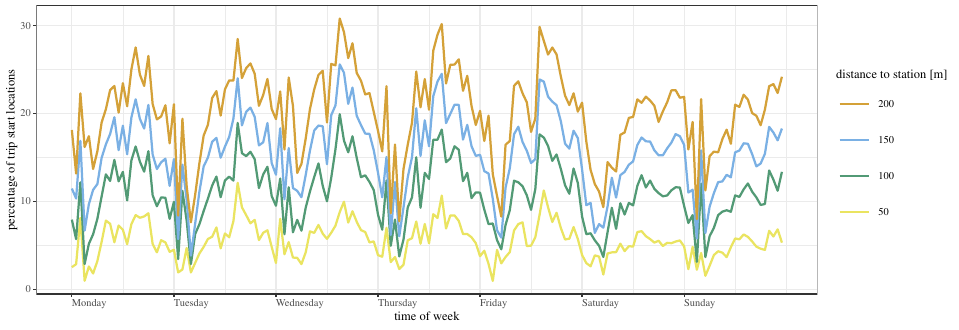}
	\includegraphics{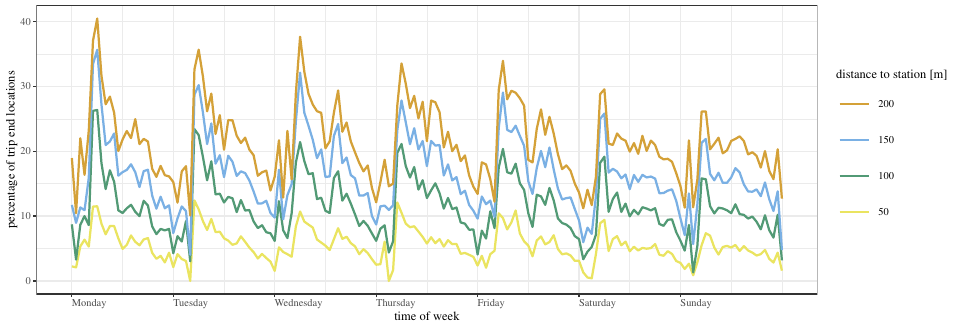}
	\caption{Percentage of trips starting (top panel) or ending (bottom panel) close to a rapid transit station. Trips that end or start close to a station exit were identified, based on the list of stations in service at the time of data collection and the location of station exits obtained from the DataMall interface~\cite{datamall}. We display results aggregated by hour over the course of the one week long dataset. Multiple threshold distances were used to compensate for uncertainties in GPS data and variability in typical bike parking locations. It is noticeable that there is a large peak in the morning considering trip destinations and a lesser peak in the afternoon considering trip origins. This suggest that these are commuting trips that use shared bikes to access the rapid transit network.}
	\label{train_dist}
\end{figure*}

\begin{figure*}
	\centering
	\includegraphics{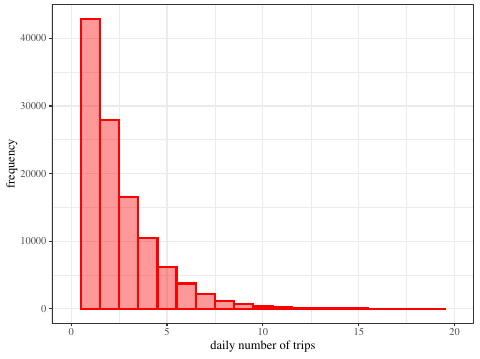}
	\includegraphics{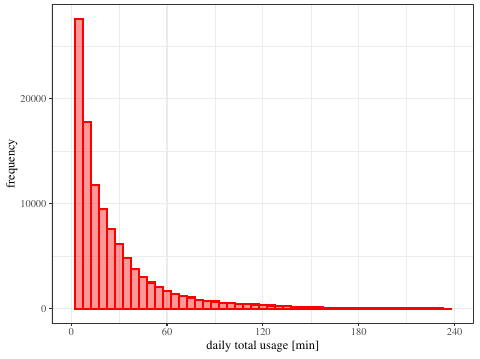}
	\includegraphics{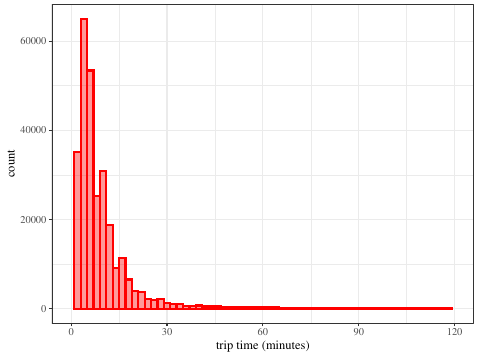}
	\includegraphics{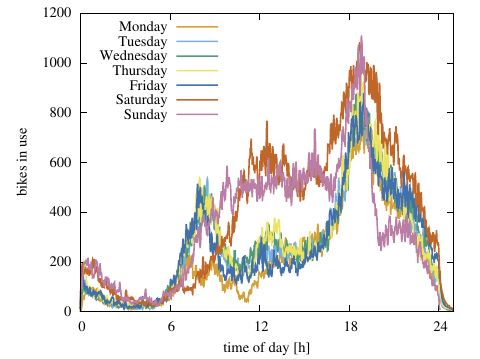}
	\caption{Basic statistics of bike usage. Top left: distribution of the number of daily trips per bike. Only active bikes (i.e.~having at least one trip that day) are considered. Most bikes are used only a few times. Top right: distribution of total time bikes are used during the day. Bottom left: distribution of trip durations. We see that most trips are shorter than 30 minutes. Bottom right: number of bikes simultaneously in use during the day (i.e.~number of trips happening simultaneously). We see that usage is highest during the evening. Also, weekdays and weekend have a distinctive pattern, with morning and evening peaks being dominant for weekdays, and a more even usage during the day for weekends.}
	\label{trip_times}
\end{figure*}

	Data was originally collected by regularly querying the public interface of a shared bike operator for all available vehicles~\cite{Shen2018,Xu2019}. After identifying trips, we filter out excessively short and long trips; the former might be the result of inaccurate GPS measurements, while the latter can correspond to the operator removing the bike for maintenance. This way, we have a total of 284,100 trips over the course of the week. We show basic statistics of trips and bike usage in our dataset in Fig.~\ref{trip_times}. We display a detailed analysis of the pattern of trips starting or ending close to rapid transit stations in Fig.~\ref{train_dist}.
	
	As the location of the bikes during the trips are not reported, we assign probable routes; we achieve this by obtaining a representation of possible paths from OpenStreetMap~\cite{haklay2008openstreetmap}, finding the shortest path for each trip and assuming it is the route taken. Currently, the use of PMDs in Singapore on sidewalks and roads is forbidden; they are only allowed to be used on cycle paths, while previously, their use on sidewalks was allowed. Worldwide, with the increasing popularity of electric scooters and PMDs, cities are adopting different regulations, with use on sidewalks and cycling routes being prominent~\cite{Herrman2019,Anderson-Hall2019}. It is uncertain what regulations will apply to SRSPMDs; in the current work, we use the assumption that they can use both sidewalks and cycling paths.
	
	After assigning shortest paths to each trip, we calculate average travel speeds and filter out trips that have an average speed above $30 \mathrm{km}/\mathrm{h}$. One probable explanation for having such trips is that the path network obtained from OpenStreetMap is incomplete, thus for some trips, our estimated ``shortest'' path is still longer than the real route taken by the user. After these processing and filtering steps, we have a total of 278,826 trips left made by 32,782 unique bikes (identified by the 9 digit unique ID for each bike reported in the dataset).

\subsection{Variants of the online model}

\begin{algorithm}[th!]
	\begin{algorithmic}[1]
		\State N: OSM path network
		\State T: $\{ t : (n_{st}, d_{st}, T_{st}, n_{et}, d_{et}, T_{et}, T_{Dt} = 0, s_t = \textrm{False}) \}$
		\State $v_R$: vehicle relocation speed
		\State $t_B$: size of batches
		\State $t = 0$ current time
		\State V: $\{ (n, t = 0) \}$ set of available vehicles (random initial distribution)
		\While{$|$T$|$ > 0}
			\State E $ = \{ t \in $T$, \, T_{et} + T_{Dt} \in [t, t + t_B), s_t = \textrm{True} \} $ %]$
			\State S $ = \{ t \in $T$, \, T_{st} < t + t_B, s_t = \textrm{False} \} $
			\ForAll{$t \in $ E}
				\Comment Process trip end events
				\State Insert(V, $(n_{et}, t_v = T_{et} + d_{et} / v_R + T_{Dt})$)
				\State Remove(T, $t$)
			\EndFor
			\State $P_T = \{ \}$ \Comment match candidates
			\ForAll{$t \in $ S} \Comment process trip start events
				\ForAll{$v \in $ V}
					\State $t_s^* = \max(t_v, t) + (d(n_v, n_{st}) + d_{st}) / v_R$
					\State Insert($P_T$, $(v, t, t_s^* - T_{st})$)
				\EndFor
			\EndFor
			
			\State $R_T = $ MaximumWeightedMatch($P_T$)
			\ForAll{$(v, t, d) \in R_T$}
				\State Remove(V, $v$)
				\State $T_{Dt} = d$
				\State $s_t = \textrm{True}$
			\EndFor
			\State $t = t + t_B$
		\EndWhile
		\State Result: T, the set of trips updated with waiting times
	\end{algorithmic}
	\caption{Variation of Algorithm~\ref{alg:online}, without limiting maximum waiting times, and using a fixed set of vehicles that are distributed randomly at the beginning instead of adding more vehicles during the day. The input is the set of trips to be served (T), the path network (N) and the parameters for vehicle relocation speed ($v_R$) and batch size ($t_B$). The result is the set of trips, T, augmented with actual waiting times for each trip. This can be used to estimate average waiting times, and the ratio of trips served under a given $t_W$ threshold.}
	\label{alg:online_nomax}
\end{algorithm}

\begin{algorithm}[th!]
	\begin{algorithmic}[1]
		\State N: OSM path network
		\State T: $\{ t : (n_{st}, d_{st}, T_{st}, n_{et}, d_{et}, T_{et}, T_{Dt} = 0) \}$ all trips
		\State $v_R$: vehicle relocation speed
		\State $t_B$: size of batches
		\State $T_{LA}$: size of look-ahead window
		\State $t_W$: maximum wait time
		\State $t = 0$ current time
		\State V: $\{ \}$ set of available vehicles (initially empty)
		\While{$|$T$|$ > 0}
			\State E $ = \{ t \in $T$, T_{et} + T_{Dt} \in [t, t + t_{LA}) \} $ %]$
			\State S $ = \{ t \in $T$, T_{st} \in [t, t + t_{LA}) \} $ %]$
			
			\State $P_T = \{ \}$ \Comment match candidates among trips
			\ForAll{$s \in $ S, $e \in $ E}
				\State $t_s^* = t_{ee} + (d(n_{ee}, n_{ss}) + d_{ss} + d_{ee}) / v_R$
				\If{$t_s^* < T_{ss} + t_W$}
					\State Insert($P_T$, $(e, s, t_s^* - T_{ss})$)
				\EndIf
			\EndFor
			
			\State $R_t = $ MaximumWeightedMatch($P_T$)
			\ForAll{$(e, s, t_D) \in R_T$, $T_{ee} < t + t_B$}
				\State Remove(T, $e$)
				\State Remove(E, $e$)
				\State Remove(S, $s$)
				\State $T_{Ds} = t_D$
			\EndFor
			
			\State $P_T = \{ \}$ \Comment match trips with free vehicles
			\ForAll{$s \in $ S, $T_{ss} < t + t_B$}
				\ForAll{$v \in $ V}
					\State $t_s^* = \max(t_v, t) + (d(n_v, n_{ss}) + d_{ss}) / v_R$
					\If{$t_s^* < T_{ss} + t_W$}
						\State Insert($P_T$, $(v, s, t_1 - T_{ss})$)
					\EndIf
				\EndFor
			\EndFor
			
			\State $R_T = $ MaximumWeightedMatch($P_T$)
			\ForAll{$(v, s, d) \in R_T$}
					\State Remove(V, $v$)
					\State Remove(S, $s$)
					\State $T_{Ds} = d$
			\EndFor
			
			\ForAll{$t \in $ S, $T_{st} < t + t_B$}
				\State A new vehicle is added at $n_{st}$ to serve $t$
			\EndFor
			
			\ForAll{$e \in $ E, $T_{ee} < t + t_B$}
				\State Insert(V, $(n_{ee}, t_v = T_{ee} + d_{ee} / v_R + T_{De})$)
				\State Remove(T, $e$)
			\EndFor
			\State $t = t + t_B$
		\EndWhile
		\State Result: $|$V$|$, the number of total vehicles used
	\end{algorithmic}
	\caption{Variation of Algorithm~\ref{alg:online}, using a longer look-ahead time window $T_{LA}$ to perform matching.}
	\label{alg:online_tw}
\end{algorithm}

Two variants of Algorithm~\ref{alg:online} are used along with the main version. Algorithm~\ref{alg:online_nomax} does not limit maximum waiting times, but also uses a fixed set of vehicles instead of dynamically increasing the number of vehicles while processing the trips. In accordance with this, the main difference is the lack of the conditional when matching vehicles to trip starts (line~18) and that the set of trip start events (S) can include unserved trips from previous batches. Also, care needs to be taken that only yet unserved trips are included in S and only already served trips are included in E. To ensure this, the extra variable $s_t$ for each trip keeps track if it was already served.

Algorithm~\ref{alg:online_tw} shows the variant that includes longer $T_{LA}$ look-ahead windows for trips. The main difference is that in each batch, all trip start and events in a larger time window ($[t, t + T_{LA})$) are included. Using these, at first, end and start events are matched directly, similarly to the oracle model. However, in accordance with the batched nature of this method, only end events in the smaller time window ($[t, t + t_B)$) are actually processed. Any remaining start events in this shorter time window are then matched to idle vehicles and finally served by newly created vehicles. Any remaining end events in this time window are processed by adding the vehicles to the V set of available vehicles.

\subsection{Definition of the path upgrade problem}

We start with a set of trips $T$ that are made by SRSPMDs, including trips made by users and repositioning trips. Our goal is to select path upgrades in a way that has the maximum impact, e.g.~maximizes travel on upgraded paths or results in maximum decrease in fleet size while serving the same set of trips. This goal is subject to a constraint on the available resources for building and upgrading paths.

The current paths available to cyclists and PMDs as obtained from OpenStreetMap form an undirected graph $G = \{V,E\}$. Each edge $e \in E$ has an associated length $l_{e}$. These are obtained as the set of all sidewalks and cycle paths. For simplicity, we assume that new, dedicated PMD paths can be built only in parallel with any of the edges present in $G$. We assume that potential upgrades to the network are carried out by edge granularity, i.e.~each edge is either upgraded or not, represented by the binary choice variable $u_{e} = 0,1$. Note that we do not consider the possibility to add edges as we assume that the path network is already sufficiently dense. In the following, we refer to a \emph{configuration} of upgrades as an $|E|$ length binary vector $\vec{u}$ which determines for all edges whether they are upgraded or not. It is clear that for a network of $N \equiv |E|$ edges, there is a total of $2^N$ distinct configurations of network upgrades. Thus, enumerating all of them is not feasible for any realistic network size.

We can formally define the path upgrade problem as the following. We define the (scalar) functions $C(\vec{u})$ and $B(\vec{u})$ as the cost and benefit of carrying out a certain upgrade respectively. We would like to carry out a network upgrade with maximal benefit subject to cost constraints:

\begin{equation}
	\textrm{find } \vec{u} \textrm{ s.t. } B(\vec{u}) \textrm{ maximal, subject to } C(\vec{u}) \leq C_\mathrm{max}
	\label{eq:opt}
\end{equation}

In this formulation, we can use $C_\mathrm{max}$ as a parameter which limits the process; repeating the solution for different values of $C_\mathrm{max}$ allows us to define $B(C)$ which determines how the maximal benefit of upgrading paths changes as a function of invested cost. To carry out this optimization, we need to define the $B$ and $C$ functions. In the current analysis, we assume that the cost of performing upgrades is simply proportional to the total length of upgraded paths, i.e.~$C(\vec{u}) = \sum_{e \in E} u_{e} l_{e}$.

In general, $B(\vec{u})$ can be a complex function of path configuration, that may not be computable analytically, but only with simulations. Essentially, for any upgrade configuration $\vec{u}$, the ideal path choices will change according to the upgrades carried out. Furthermore, the definition of $B$ can differ based on which stakeholder's perspective is considered: e.g.~for an operator, it can be the reduction in fleet size achievable due to increased speed; for users, it can be the decrease in travel and waiting times; for the city, in can be the decrease in potential conflicts between path users. In the current work, we adopt the simple measure where $B$ is defined as the total relative length of travel on upgraded paths:

\begin{equation}
	B(\vec{u}) = \frac{ \sum_{e \in E} u_{e} l_{e} n_{e} }{ \sum_{e \in E} l_{e} n_{e} } = \frac{ \sum_{t \in T} \sum_{e \in e_t} u_{e} l_{e} }{ \sum_{t \in T} \sum_{e \in e_t} l_{e} }
	\label{eq:bsimple}
\end{equation}

Here we denote the number of distinct trips that traverse edge $e$ by $n_{e}$. Alternatively, we can calculate the same measure by summing up the length of upgraded path segments for each trip separately: doing this on the right hand side of Eq.~\ref{eq:bsimple}, by $e_t$, we denote the set of edges traversed by trip $t$. Even in this case, $B(\vec{u})$ will depend on the solution of the routing and vehicle assignment problems given the configuration of upgraded paths $\vec{u}$. Formally, we note that the set of trips $T$ and the set of edges for each trip $t$ will depend on $\vec{u}$ since an upgraded path configuration can result in alternative routes becoming preferable and also in more repositioning trips becoming possible between trips.

We present two approximations for $B(\vec{u})$:
\begin{itemize}
	\item $B^{*}$ is an approximation assuming that neither the routing of trips nor the assignment of vehicles changes as a result of path upgrades.
	\item $B^{T}$ is an approximation assuming that the set of trips does not change, but each trip is rerouted to find an ideal path given the upgraded network.
\end{itemize}

Essentially, we have
\begin{gather}
	B^{*}(\vec{u}) = \frac{ \sum_{t \in T^*} \sum_{e \in e_t^*} u_{e} l_{e} }{ \sum_{t \in T^*} \sum_{e \in e_t^*} l_{e} }
	\label{eq:bapprox} \\
	B^{T}(\vec{u}) = \frac{ \sum_{t \in T^*} \sum_{e \in e_t} u_{e} l_{e} }{ \sum_{t \in T^*} \sum_{e \in e_t} l_{e} }
	\label{eq:bapproxt}
\end{gather}
where $T^*$ is the original set of trips (with vehicle assignments calculated without considering path upgrades) and for each trip, $e_t^*$ is the original set of edges that it traverses. $B^*$ can then be easily calculated based on the solutions of the original routing and vehicle assignment problem given a configuration of path upgrades $\vec{u}$, while calculating $B^T$ requires solving the routing problem again after considering the updates. It follows that for any given set of updates $\vec{u}$, $B^T(\vec{u}) \geq B^*(\vec{u})$, since any path that allows faster travel will have a higher ratio of upgraded path lengths. It is less clear if after solving the vehicle assignment problem on the upgraded paths, the value for $B(\vec{u})$ is increased as well. We conjecture that this is true, giving:
\begin{equation}
	B(\vec{u}) \geq B^T(\vec{u}) \geq B^*(\vec{u}) \quad \quad \forall \vec{u}
\end{equation}

The approximation $B^*(C)$ can be easily computed as a function of maximum upgrade costs. Given that in this case, the denominator in Eq.~\ref{eq:bsimple} is constant, we note that $B^*$ is proportional to a linear combination of $l_e n_e$ values. We can maximize $B^*$ by including the terms with the highest $n_e$ values until the limit $C$ is reached. While for a general value of $C$, calculating $B^*$ this way presents the knapsack problem, we can calculate a discrete number of ideal solutions by successively adding edges to the upgraded set by decreasing rank. For a given solution of the routing and vehicle assignment problem, we define an ordering of edges $e_1, e_2, \ldots e_N$ such that $n_{e_{j+1}} \leq n_{e_j}$, $\forall j$. Next, we define a series of upgrade choices, $\vec{u}_i$ ($i = 1,2 \ldots N$), where $u_{ij}$ decides whether edge $e_j$ is upgraded according to the same ordering:
\begin{equation}
	u_{ij} = \left \{ \begin{array}{ll}
		1 & j \leq i \\
		0 & \textrm{otherwise}
	\end{array} \right .
	\label{eq:ui}
\end{equation}
and then a series of cost and benefit values $C_i \equiv C(\vec{u}_i)$, $B_i^* \equiv B^* (\vec{u}_i)$. It follows that $B^*_i$ is the ideal solution of the problem defined in Eq.~\ref{eq:opt} for the approximate benefit function defined in Eq.~\ref{eq:bapprox} and the constraint $C_\mathrm{max} \equiv C_i$. Exploiting that $B^*(C)$ is a nondecreasing function, we can then approximate it with the step function:

\begin{equation}
	B^*(C) \equiv B^*_{i+1} \quad \textrm{if} \quad C \in [ C_i, C_{i+1} ) % ]
\end{equation}
(using the natural extensions $C_0 \equiv 0$, $C_{N+1} \equiv C_N$ and $B^*_{N+1} \equiv B^*_N$). We display this function in Fig.~\ref{improvements1} in the Supplementary Material, separately for the set of original trips and repositioning trips, along with the values of $B^T (\vec{u}_i)$, calculated for a subset of points, using the same $\vec{u}_i$ upgrade configuration that maximizes $B^*$ in those points.

In the analysis presented in Section~\ref{sec:improvements} in the main text and in Figs.~\ref{improvements1} -- \ref{improved_online_ntrips1} in the Supplementary Material, we used the approximate form of $B^*(\vec{u})$ as described in Eq.~\ref{eq:bapprox}. We selected path segments to upgrade based on the scheme presented in Eq.~\ref{eq:ui}. In this way, our results present a lower bound on the benefits of path upgrades when considering fleet size and vehicle utilization. Using a benefit function that directly evaluates savings in fleet size could allow more optimizations, although it would present tremendous computational challenges, as it would result in an optimization problem where the objective function can only be computed via time-consuming simulations.

\begin{table*}
	\centering
	% {\huge \bf Supplementary Material} \\[5ex]
	\begin{tabular}{r||r|r|r|r|r|r}
		& \multicolumn{6}{c}{avg. trips / vehicle}  \\
		day & \multicolumn{2}{c|}{shared bikes} & \multicolumn{4}{c}{SRSPMDs} \\
		 & current & ideal & 1 km/h & 2.5 km/h & 5 km/h & 10 km/h  \\ \hline
		Monday &    2.30	& 2.84	& 8.80	& 12.62	& 16.43	& 20.69 \\ \hline
		Tuesday &   2.44	& 3.05	& 9.69	& 14.05	& 18.21	& 23.08 \\ \hline
		Wednesday & 2.48	& 3.09	& 10.48	& 15.34	& 20.00	& 25.30 \\ \hline
		Thursday &  2.48	& 3.07	& 10.20	& 14.85	& 19.19	& 23.67 \\ \hline
		Friday &    2.37	& 2.96	& 10.34	& 14.90	& 19.54	& 24.61 \\ \hline
		Saturday &  2.75	& 3.55	& 12.12	& 17.32	& 21.98	& 27.32 \\ \hline
		Sunday &    2.63	& 3.39	& 11.34	& 15.77	& 19.62	& 23.80 \\
	\end{tabular}
	\caption{Main results for average vehicle utilization (trips / vehicle).}
	\label{usage_trips_tab}
\end{table*}

\begin{table*}
	\centering
	\begin{tabular}{r|r||r|r|r||r|r|r|r||r}
		& & \multicolumn{7}{c||}{fleet size} & \multirow{3}{*}{\makecell{maximum\\ utilization}} \\
		day & trips & \multicolumn{3}{c||}{shared bikes} & \multicolumn{4}{|c||}{SRSPMDs} & \\ 
		 & & available & used & ideal & 1 km/h & 2.5 km/h & 5 km/h & 10 km/h & \\ \hline
		Monday &    31058 & 35350 & 13512 & 10944 & 3531 & 2461 & 1890 & 1501 & 800  \\ \hline 
		Tuesday &   38030 & 35535 & 15556 & 12488 & 3926 & 2706 & 2088 & 1648 & 859  \\ \hline
		Wednesday & 39645 & 37642 & 15984 & 12828 & 3784 & 2584 & 1982 & 1567 & 876  \\ \hline
		Thursday &  39537 & 37885 & 15929 & 12878 & 3878 & 2663 & 2060 & 1670 & 966  \\ \hline
		Friday &    37143 & 38229 & 15702 & 12560 & 3592 & 2492 & 1901 & 1509 & 878  \\ \hline
		Saturday &  48747 & 38128 & 17743 & 13729 & 4021 & 2814 & 2218 & 1784 & 1083 \\ \hline
		Sunday &    44666 & 38150 & 16956 & 13192 & 3938 & 2832 & 2277 & 1877 & 1110 \\
	\end{tabular}
	\caption{Main results for fleet size. Fleet sizes for SRSPMDs with different assumed average relocation speed are compared to current fleet size of shared bikes, minimum achievable fleet size of shared bikes with users willing to walk up to $100\,\mathrm{m}$, and to the fleet size corresponding to the maximum number of bikes simultaneously in use each day.}
	\label{fleet_tab}
\end{table*}

\begin{table*}
	\centering
	\begin{tabular}{r||r|r|r|r|r|r}
		& \multicolumn{6}{c}{avg. time used / vehicle [min]} \\
		day & \multicolumn{2}{c|}{shared bikes} & \multicolumn{4}{c}{SRSPMDs} \\
		 & current & ideal & 1 km/h & 2.5 km/h & 5 km/h & 10 km/h  \\ \hline
		Monday &    23.21	& 28.66	& 88.83	 & 127.45	& 165.95	& 208.96 \\ \hline
		Tuesday &   24.09	& 30.01	& 95.47	 & 138.51	& 179.50	& 227.43 \\ \hline
		Wednesday & 24.37	& 30.36	& 102.92 & 	150.72	& 196.49	& 248.53 \\ \hline
		Thursday &  24.80	& 30.68	& 101.88 & 	148.36	& 191.79	& 236.58 \\ \hline
		Friday &    23.76	& 29.71	& 103.88 & 	149.73	& 196.28	& 247.27 \\ \hline
		Saturday &  32.40	& 41.88	& 142.98 & 	204.31	& 259.21	& 322.27 \\ \hline
		Sunday &    29.40	& 37.78	& 126.58 & 	176.01	& 218.91	& 265.56 \\
	\end{tabular}
	\caption{Main results for average vehicle utilization (time used / vehicle).}
	\label{usage_time_tab}
\end{table*}

\begin{table*}
	\centering
	\begin{tabular}{l||r|r|r|r|r}
		\multirow{3}{*}{day} & \multicolumn{5}{c}{total distance traveled [km]} \\
		& \multirow{2}{*}{\makecell{original\\ trips}} & \multicolumn{4}{c}{connecting trips}  \\
		& &			1 km/h &	2.5 km/h &	5 km/h &	10 km/h \\ \hline \hline
		Monday	    & 26171 &	7527  &	10064 &	12524 &	15270 \\ \hline
		Tuesday	    & 31971 &	9653  &	13057 &	15718 &	18759 \\ \hline
		Wednesday	& 33219 &	10111 &	13683 &	16739 &	20281 \\ \hline
		Thursday	& 32780 &	9846  &	13207 &	16176 &	19188 \\ \hline
		Friday	    & 31473 &	10318 &	13491 &	16617 &	19962 \\ \hline
		Saturday	& 42806 &	12225 &	16110 &	19522 &	23555 \\ \hline
		Sunday	    & 38530 &	11123 &	14175 &	16477 &	19231
	\end{tabular}
	\caption{Total distance traveled by users during the original trips and by SRSPMDs during connecting trips with different assumed relocation speed. We see that the distance traveled by the vehicles without a user (in connecting trips) can exceed 50\% of the distance traveled with a user.}
	\label{tab_distances1}
\end{table*}

\begin{table*}
	\centering
	\begin{tabular}{l||r|r|r|r|r}
		\multirow{3}{*}{day} & \multicolumn{5}{c}{avg. distance traveled / vehicle [km]} \\
		& \multirow{2}{*}{\parbox{1.2cm}{original fleet}} & \multicolumn{4}{c}{SRSPMD fleet}  \\
		& &			1 km/h &	2.5 km/h &	5 km/h &	10 km/h \\ \hline \hline
		Monday	    & 1.94 &	9.54 &	14.72 &	20.47 &	27.61 \\ \hline
		Tuesday	    & 2.06 &	10.60 &	16.64 &	22.84 &	30.78 \\ \hline
		Wednesday	& 2.08 &	11.45 &	18.15 &	25.21 &	34.14 \\ \hline
		Thursday	& 2.06 &	10.99 &	17.27 &	23.77 &	31.12 \\ \hline
		Friday	    & 2.00 &	11.63 &	18.04 &	25.30 &	34.09 \\ \hline
		Saturday	& 2.41 &	13.69 &	20.94 &	28.10 &	37.20 \\ \hline
		Sunday	    & 2.27 &	12.61 &	18.61 &	24.16 &	30.77
	\end{tabular}
	\caption{Average distance traveled by each vehicle in the fleet.}
	\label{tab_distances2}
\end{table*}

\begin{figure*}
	\includegraphics{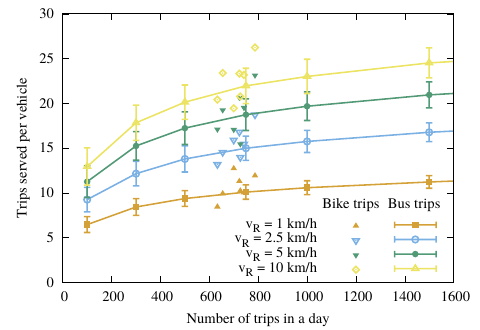}
	\includegraphics{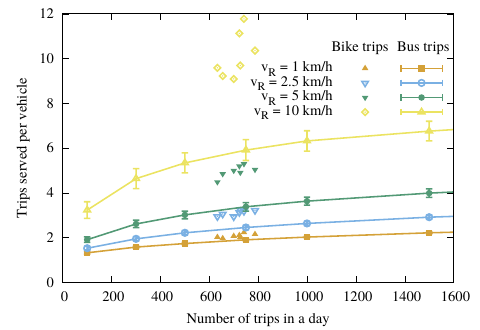}
	\caption{Detailed results for serving bus trips by SRSPMDs in the Toa Payoh area. Left: oracle model; right: online model. In these figures, dots show results calculated based on the bikeshare dataset, limited to trips happening in the Toa Payoh area. These are in good agreement with results based on bus trips data in the oracle model, while there is a significant difference for the online model.}
	\label{bustrips_2}
\end{figure*}

\begin{figure*}
	\centering
	\includegraphics{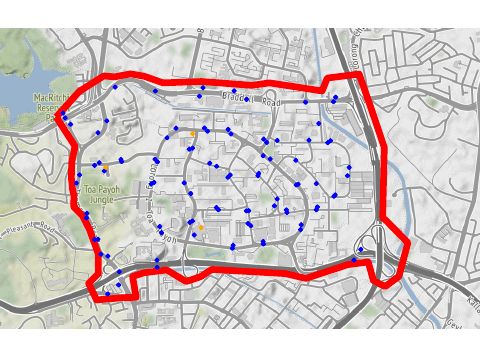}
	\includegraphics{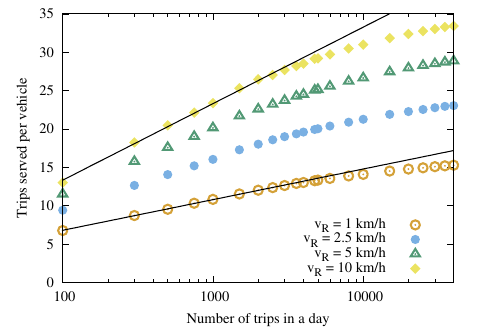}
		
	\caption{Left: Study area for bus trip data. Bus trips within the area marked by the red border are used in our analysis. Blue dots represent bus stops considered in this study, while the orange dots are the three MRT stations in this area (Toa Payoh and Braddel on the North-South Line and Caldecott on the Circle Line). Note that we excluded three bus stops on the north side of Braddel Road (the northern border of the study area) as it is unclear if crossing the road from there would be possible for SRSPMDs. Right: daily average utilization of SRSPMDs as a function of daily number of trips with a logarithmic $x$-axis. We see that vehicle utilization can be well modeled to grow logarithmically as a function of the daily number of trips until about 5000 trips / day, when a saturation effect becomes significant. The black lines are fits of logarithmic growth for the $v_R = 1\,\mathrm{km}/\mathrm{h}$ and $v_R = 10\,\mathrm{km}/\mathrm{h}$ cases.}
	\label{toapayoh}
\end{figure*}

\begin{figure*}
	\centering
	\includegraphics{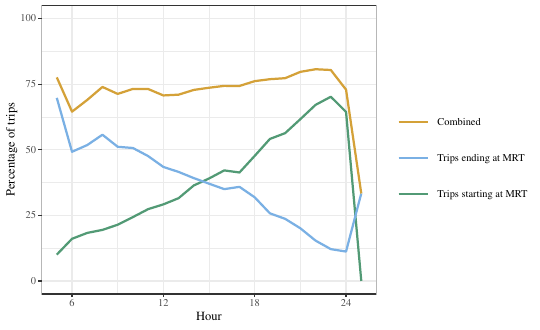}
	\caption{Ratio of bus trips starting or ending at an MRT station among all bus trips taken inside the Toa Payoh area. The green line further shows combined ratios, i.e.~the ratio of all trips that either start or end at a station. Very few trips have both their start and end close to a station, thus this combined figure is very close to the sum of the separate ratios. We see that the ratio of trips ending at a station is highest in the morning and gradually decreases during the day, while the ratio of trips starting at a station displays an opposite trend. This suggests that a large amount of these trips are made by commuters who live in this area and use the buses to access the MRT. The overall ratio for the whole day is $37\%$ and $39.7\%$ for trip starts and ends respectively, with a combined total of $73.6\%$. Note that this statistic is only calculated here for the bus trips that start and end within the study area of Toa Payoh and thus are typically short trips.}
	\label{bus_mrt_ratio}
\end{figure*}

\begin{figure*}
	\centering
	\includegraphics{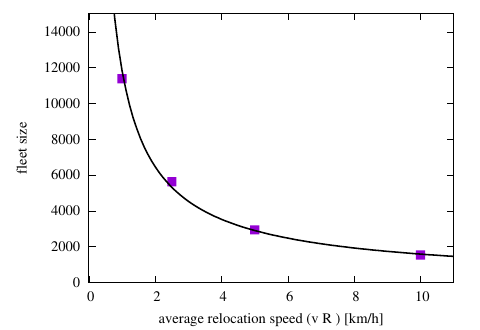}
	\includegraphics{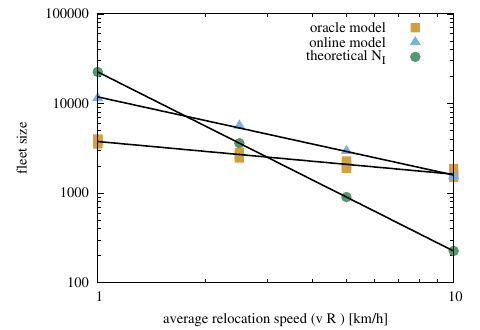}
	\caption{Estimating the relationship between average relocation speed and fleet size. Left: number of vehicles needed to serve at least 50\% of trips with up to $5\,\mathrm{min}$ waiting time in the online model. The 50\% threshold was chosen to obtain a robust estimation of fleet size scaling with vehicle speed. The fitted power-law function has an exponent of $-0.872$. Right: comparison of scaling in the online model with the oracle model and the simple estimate of $N_I$ vehicles covering all of the service area. The exponents are $-0.365$ (oracle model), $-0.872$ (online model) and $-2$ ($N_I$ estimate). For low relocation speeds, the concern for availability of vehicles (represented by $N_I$) is dominant. Real fleet sizes are smaller: in the oracle model, relocation decisions are made in advance, thus positioning vehicles close to demand is less important; in the case of the online model, we only require 50\% of trips to be served within $5\,\mathrm{min}$ waiting time, thus a smaller fleet is effective. For larger relocation speeds ($v_R \geq 5\,\mathrm{km} / \mathrm{h}$), $N_I$ quickly becomes very small compared to the fleet size determined by the vehicles performing trips or engaged in relocation movements.}
	\label{online_5min_fleet}
\end{figure*}

\begin{figure*}
	\centering
	\begin{minipage}{3.2in}
		\centering
		\includegraphics{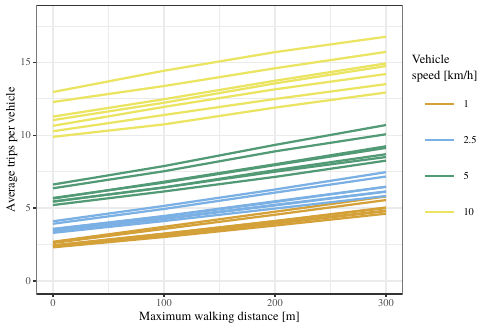}% \quad
		\captionof{figure}{Exploring the effect of passengers walking to meet vehicles on required fleet sizes in the online model. In this case, we assumed that users are willing to walk up to a maximum distance to meet a vehicle that they can use for their trip. Walking speed was assumed to be $1\, \mathrm{m} / \mathrm{s}$ ($3.6\, \mathrm{km} / \mathrm{h}$, which is notably faster than the relocation speed  of SRSPMDs at the two slower assumed cases). We used a maximum waiting time of $5$ minutes, which included any time already spent walking as well. In this figure we show changes in average vehicle utilization, where different lines correspond to different days in our dataset.}
		\label{bikeshare_online_walk}
	\end{minipage}
\quad\quad\quad
	\begin{minipage}{3.2in}
		\centering
		\includegraphics{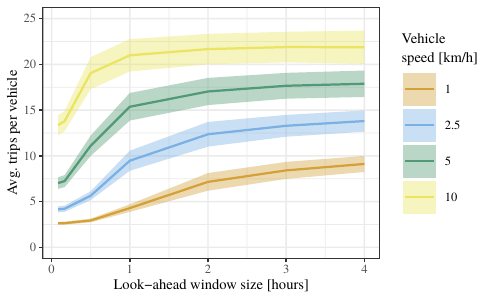}
		\captionof{figure}{Average fleet utilization as a function of look-ahead window size $T_{LA}$ in the limited oracle model. We varied $T_{LA}$ from 5 minutes to 4 hours and require trips to be served without waiting. Results are averaged over one week of data; ribbons show the standard deviation of data over a week. Separate results for each day are shown in Fig.~\ref{bikeshare_online_tw2}.}
		\label{bikeshare_online_tw1}
	\end{minipage}
\end{figure*}

\begin{figure*}
	\centering
	\begin{overpic}{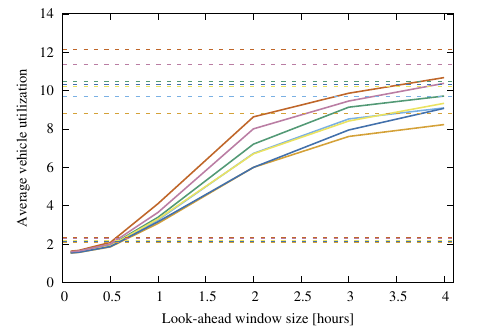}
		\put(17,68){\small $v_R = 1\,\mathrm{km}/\mathrm{h}$}
	\end{overpic}
	\begin{overpic}{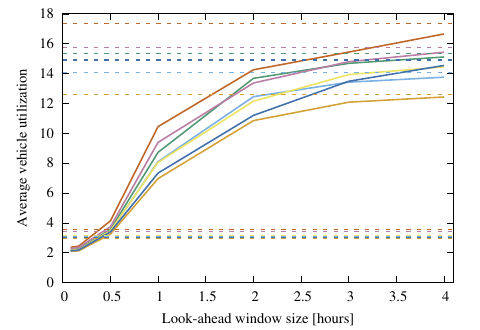}
		\put(17,68){\small $v_R = 2.5\,\mathrm{km}/\mathrm{h}$}
	\end{overpic} \\[3ex]
	\begin{overpic}{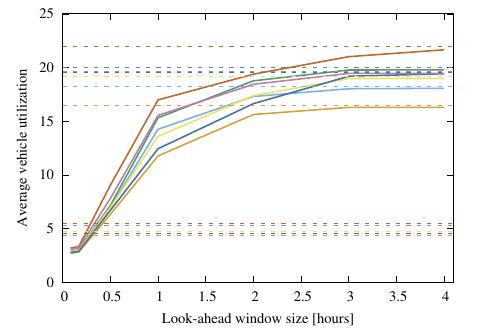}
		\put(17,68){\small $v_R = 5\,\mathrm{km}/\mathrm{h}$}
	\end{overpic}
	\begin{overpic}{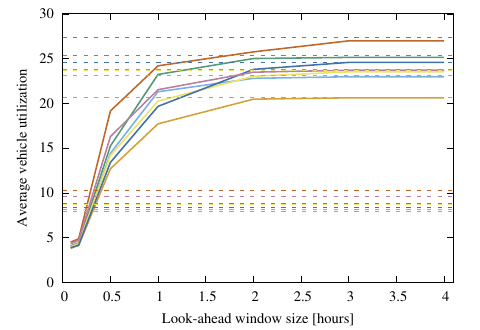}
		\put(17,68){\small $v_R = 10\,\mathrm{km}/\mathrm{h}$}
	\end{overpic}
	\includegraphics{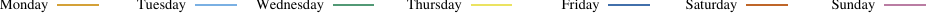}
	\caption{Improvements in average vehicle utilization in the limited oracle model as a function of $T_{LA}$, the look-ahead window size. Solid lines show the change in average vehicle utilization as $T_{LA}$ is varied from 5 minutes to 4 hours. Dashed lines show the results for the online and oracle model. We note that utilization for small $T_{LA}$ values can be worse than in the online model as we do not allow any delay for serving trips (the results for the online model were obtained with allowing a maximum of $t_{w} = 5\,\mathrm{min}$ waiting time for passengers before the start of trips). Results for larger $T_{LA}$ window sizes approach those obtained in the oracle model.}
	\label{bikeshare_online_tw2}
\end{figure*}

\begin{figure*}
	\centering
	\includegraphics{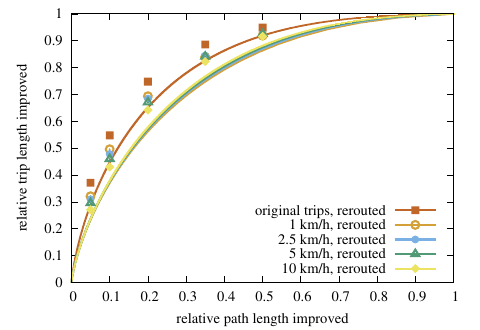}
	\caption{Quantifying the benefits of improving paths. We display with lines the ratio of cumulative travel on improved paths as a function of the ratio of total path length upgraded, under the assumption that routing and vehicle dispatching decisions do not change (i.e.~the approximation of benefits by $B^*(C)$ defined in Eq.~\ref{eq:bapprox}). We see that small improvements in the path network will affect relatively large share of total distance traveled, e.g.~upgrading 26.2\% of total length of the path network (approximately $1{,}500\,\mathrm{km}$ of paths) will improve 73.8\% of all trips by distance (approximately $175{,}000\,\mathrm{km}$ travel by bike users in one week). Results are shown separately for the original and relocation trips. Points denote the ratio of upgraded paths used after recalculating the routing of trips on the upgraded network (i.e.~the approximation $B^T$ as defined in Eq.~\ref{eq:bapproxt}). We see that these present further benefits over the simpler approximation given by $B^*$, e.g.~upgrading only 10\% of paths allows for about 55\% of trips to happen on upgraded path segments.}
	\label{improvements1}
\end{figure*}

\begin{figure*}
	\centering
	\begin{overpic}{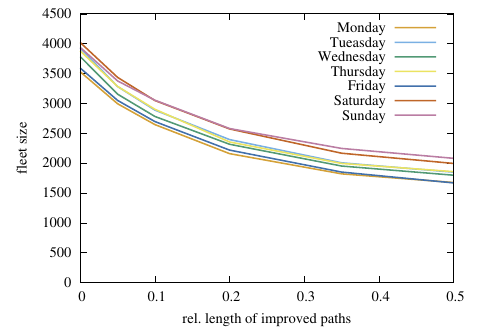}
		\put(17,68){$v_R = 1\,\mathrm{km}/\mathrm{h}$, oracle model}
	\end{overpic}
	\begin{overpic}{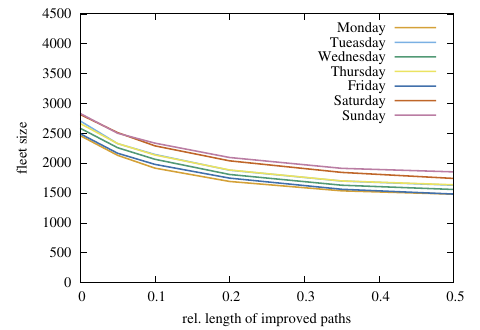}
		\put(17,68){$v_R = 2.5\,\mathrm{km}/\mathrm{h}$, oracle model}
	\end{overpic} \\[3ex]
	\begin{overpic}{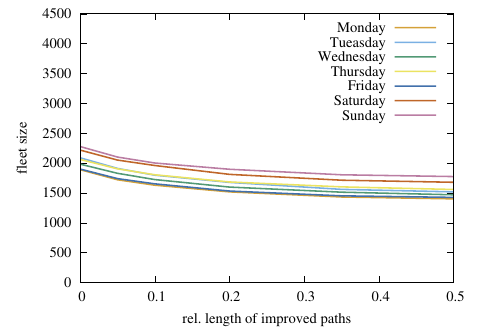}
		\put(17,68){$v_R = 5\,\mathrm{km}/\mathrm{h}$, oracle model}
	\end{overpic}
	\begin{overpic}{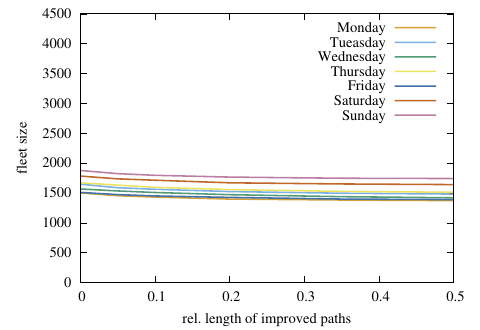}
		\put(17,68){$v_R = 10\,\mathrm{km}/\mathrm{h}$, oracle model}
	\end{overpic}
	
	\caption{Reductions in fleet size in the oracle model due to upgrading parts of the path network with original $v_R = 1\,\mathrm{km/h}$ (top left), $v_R = 2.5\,\mathrm{km/h}$ (top right), $v_R = 5\,\mathrm{km/h}$ (bottom left) and $v_R = 10\,\mathrm{km/h}$ (bottom right).}
	\label{improved_fleetsize1}
\end{figure*}

\begin{figure*}
	\centering
	\begin{overpic}{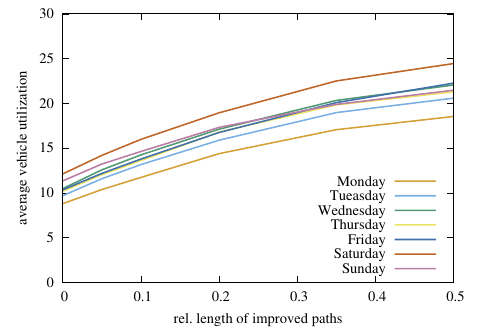}
		\put(17,68){$v_R = 1\,\mathrm{km}/\mathrm{h}$, oracle model}
	\end{overpic}
	\begin{overpic}{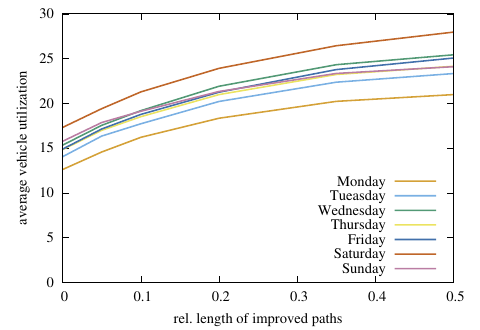}
		\put(17,68){$v_R = 2.5\,\mathrm{km}/\mathrm{h}$, oracle model}
	\end{overpic} \\[3ex]
	\begin{overpic}{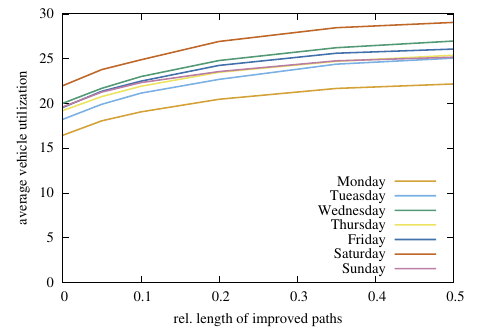}
		\put(17,68){$v_R = 5\,\mathrm{km}/\mathrm{h}$, oracle model}
	\end{overpic}
	\begin{overpic}{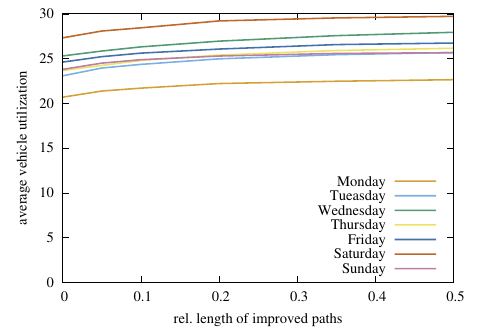}
		\put(17,68){$v_R = 10\,\mathrm{km}/\mathrm{h}$, oracle model}
	\end{overpic}
	\caption{Increase in average fleet utilization in the oracle model due to upgrading parts of the path network with original $v_R = 1\,\mathrm{km/h}$ (top left), $v_R = 2.5\,\mathrm{km/h}$ (top right), $v_R = 5\,\mathrm{km/h}$ (bottom left) and $v_R = 10\,\mathrm{km/h}$ (bottom right).}
	\label{improved_utilization1}
\end{figure*}

\begin{figure*}
	\centering
	\begin{overpic}{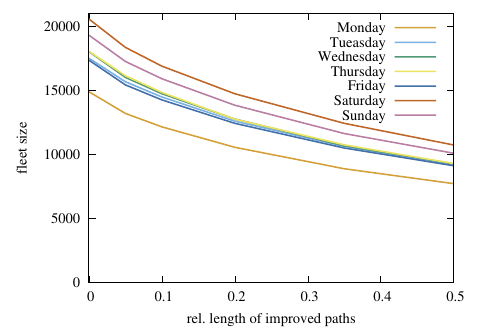}
		\put(17,68){$v_R = 1\,\mathrm{km}/\mathrm{h}$, online model}
	\end{overpic}
	\begin{overpic}{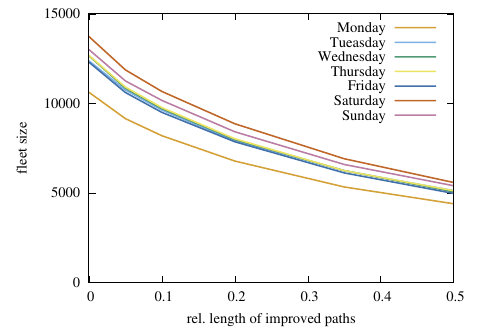}
		\put(17,68){$v_R = 2.5\,\mathrm{km}/\mathrm{h}$, online model}
	\end{overpic} \\[3ex]
	\begin{overpic}{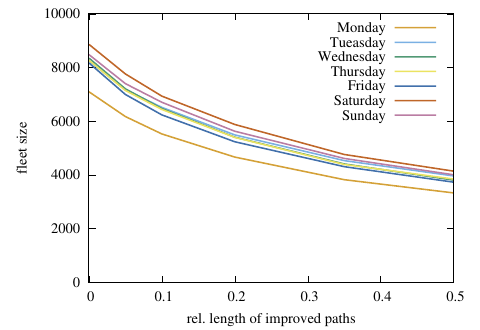}
		\put(17,68){$v_R = 5\,\mathrm{km}/\mathrm{h}$, online model}
	\end{overpic}
	\begin{overpic}{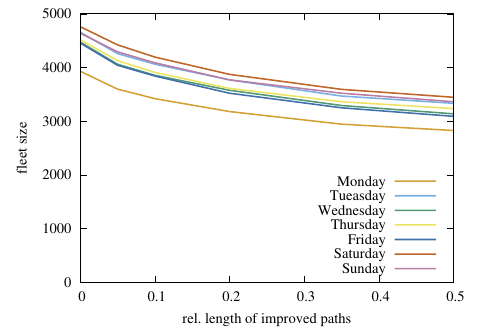}
		\put(17,68){$v_R = 10\,\mathrm{km}/\mathrm{h}$, online model}
	\end{overpic}
	\caption{Reductions in fleet size in the online model due to upgrading parts of the path network with original $v_R = 1\,\mathrm{km/h}$ (top left), $v_R = 2.5\,\mathrm{km/h}$ (top right), $v_R = 5\,\mathrm{km/h}$ (bottom left) and $v_R = 10\,\mathrm{km/h}$ (bottom right).}
	\label{online_improved_fleetsize1}
\end{figure*}

\begin{figure*}
	\centering
	\begin{overpic}{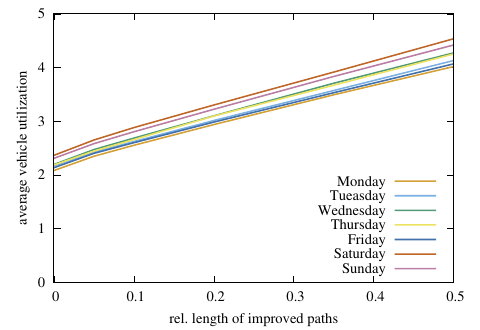}
		\put(17,68){$v_R = 1\,\mathrm{km}/\mathrm{h}$, online model}
	\end{overpic}
	\begin{overpic}{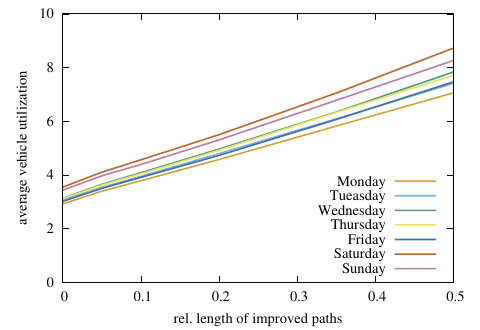}
		\put(17,68){$v_R = 2.5\,\mathrm{km}/\mathrm{h}$, online model}
	\end{overpic} \\[3ex]
	\begin{overpic}{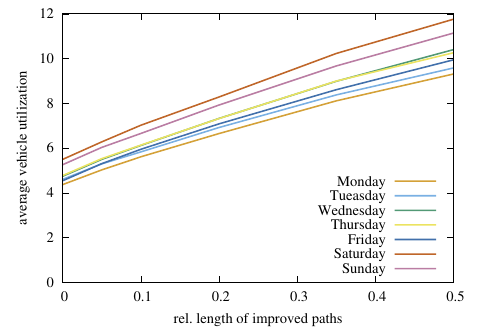}
		\put(17,68){$v_R = 5\,\mathrm{km}/\mathrm{h}$, online model}
	\end{overpic}
	\begin{overpic}{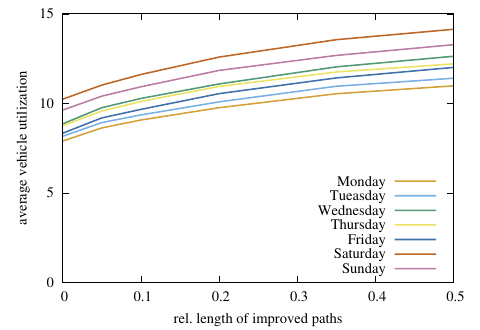}
		\put(17,68){$v_R = 10\,\mathrm{km}/\mathrm{h}$, online model}
	\end{overpic}
	\caption{Increase in average fleet utilization in the online model due to upgrading parts of the path network with original $v_R = 1\,\mathrm{km/h}$ (top left), $v_R = 2.5\,\mathrm{km/h}$ (top right), $v_R = 5\,\mathrm{km/h}$ (bottom left) and $v_R = 10\,\mathrm{km/h}$ (bottom right).}
	\label{online_improved_utilization1}
\end{figure*}

\begin{figure*}
	\centering
	\begin{overpic}{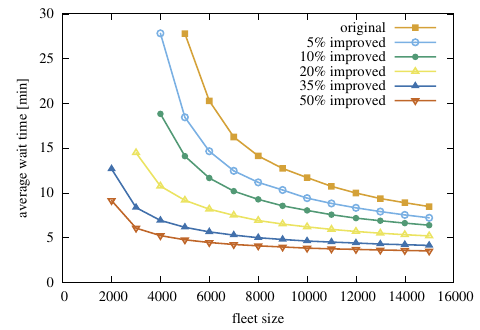}
		\put(17,68){$v_R = 1\,\mathrm{km}/\mathrm{h}$, online model}
	\end{overpic}
	\begin{overpic}{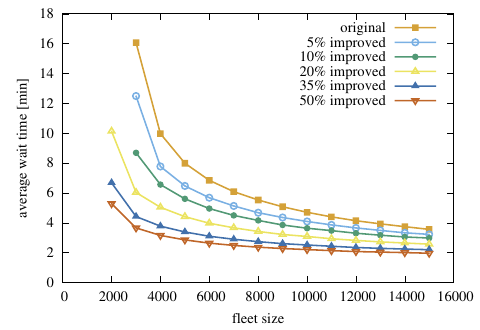}
		\put(17,68){$v_R = 2.5\,\mathrm{km}/\mathrm{h}$, online model}
	\end{overpic} \\[3ex]
	\begin{overpic}{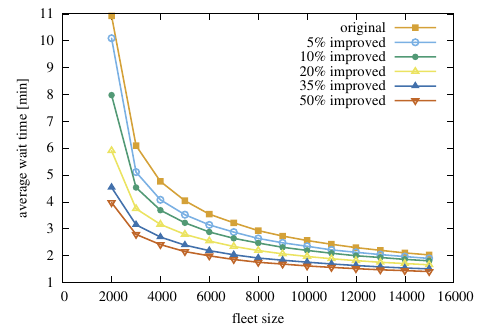}
		\put(17,68){$v_R = 5\,\mathrm{km}/\mathrm{h}$, online model}
	\end{overpic}
	\begin{overpic}{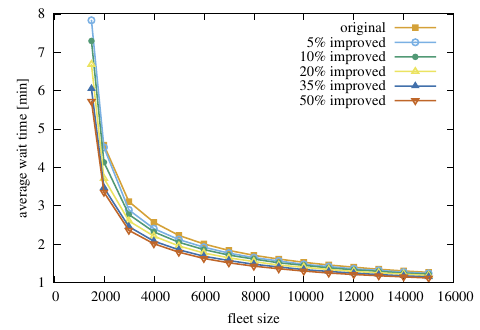}
		\put(17,68){$v_R = 10\,\mathrm{km}/\mathrm{h}$, online model}
	\end{overpic}
	\caption{Average waiting times in the online model and improvements due to infrastructure upgrades for original travel speed $v_R = 1\,\mathrm{km/h}$ (top left), $v_R = 2.5\,\mathrm{km/h}$ (top right), $v_R = 5\,\mathrm{km/h}$ (bottom left) and $v_R = 10\,\mathrm{km/h}$ (bottom right).}
	\label{improved_online_wait2}
\end{figure*}

\begin{figure*}
	\centering
	\begin{overpic}{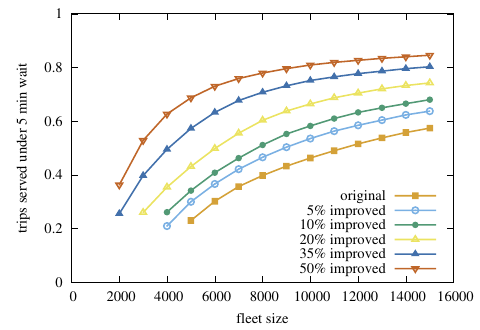}
		\put(17,68){$v_R = 1\,\mathrm{km}/\mathrm{h}$, online model}
	\end{overpic}
	\begin{overpic}{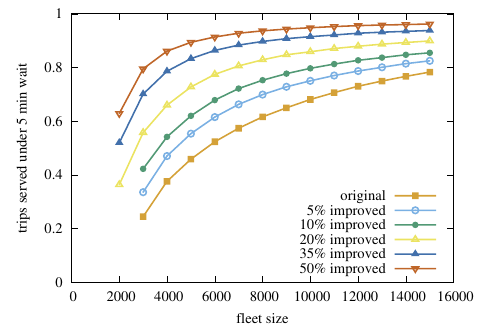}
		\put(17,68){$v_R = 2.5\,\mathrm{km}/\mathrm{h}$, online model}
	\end{overpic} \\[3ex]
	\begin{overpic}{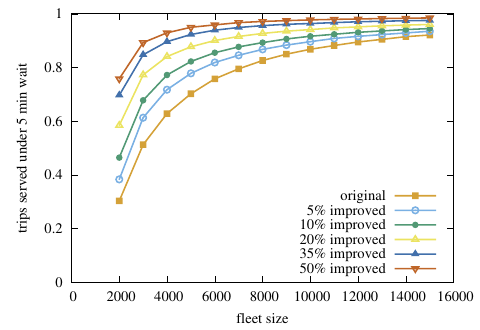}
		\put(17,68){$v_R = 5\,\mathrm{km}/\mathrm{h}$, online model}
	\end{overpic}
	\begin{overpic}{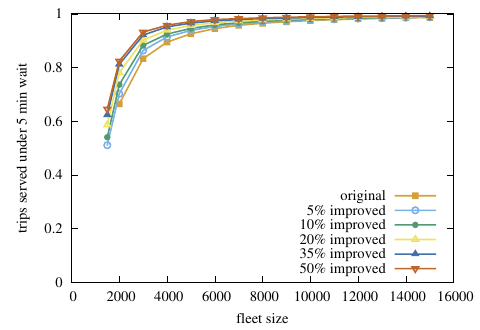}
		\put(17,68){$v_R = 10\,\mathrm{km}/\mathrm{h}$, online model}
	\end{overpic}
	\caption{Ratio of trips served within $t_w = 5\,$minutes waiting time in the online model and improvements due to infrastructure upgrades as a function of fleet size for $v_R = 1\,\mathrm{km/h}$ (top left), $v_R = 2.5\,\mathrm{km/h}$ (top right), $v_R = 5\,\mathrm{km/h}$ (bottom left) and $v_R = 10\,\mathrm{km/h}$ (bottom right).}
	\label{improved_online_ntrips1}
\end{figure*}

\begin{figure*}
\centering
%~ \includegraphics[width=0.47\textwidth]{fleetsize_cost_wait_time/1kmph_bold.png} \,
%~ \includegraphics[width=0.47\textwidth]{fleetsize_cost_wait_time/2_5kmph_bold.png} \\[3ex]
%~ \includegraphics[width=0.47\textwidth]{fleetsize_cost_wait_time/5kmph_bold.png} \,
%~ \includegraphics[width=0.47\textwidth]{fleetsize_cost_wait_time/10kmph_bold.png}
\begin{overpic}[width=0.47\textwidth]{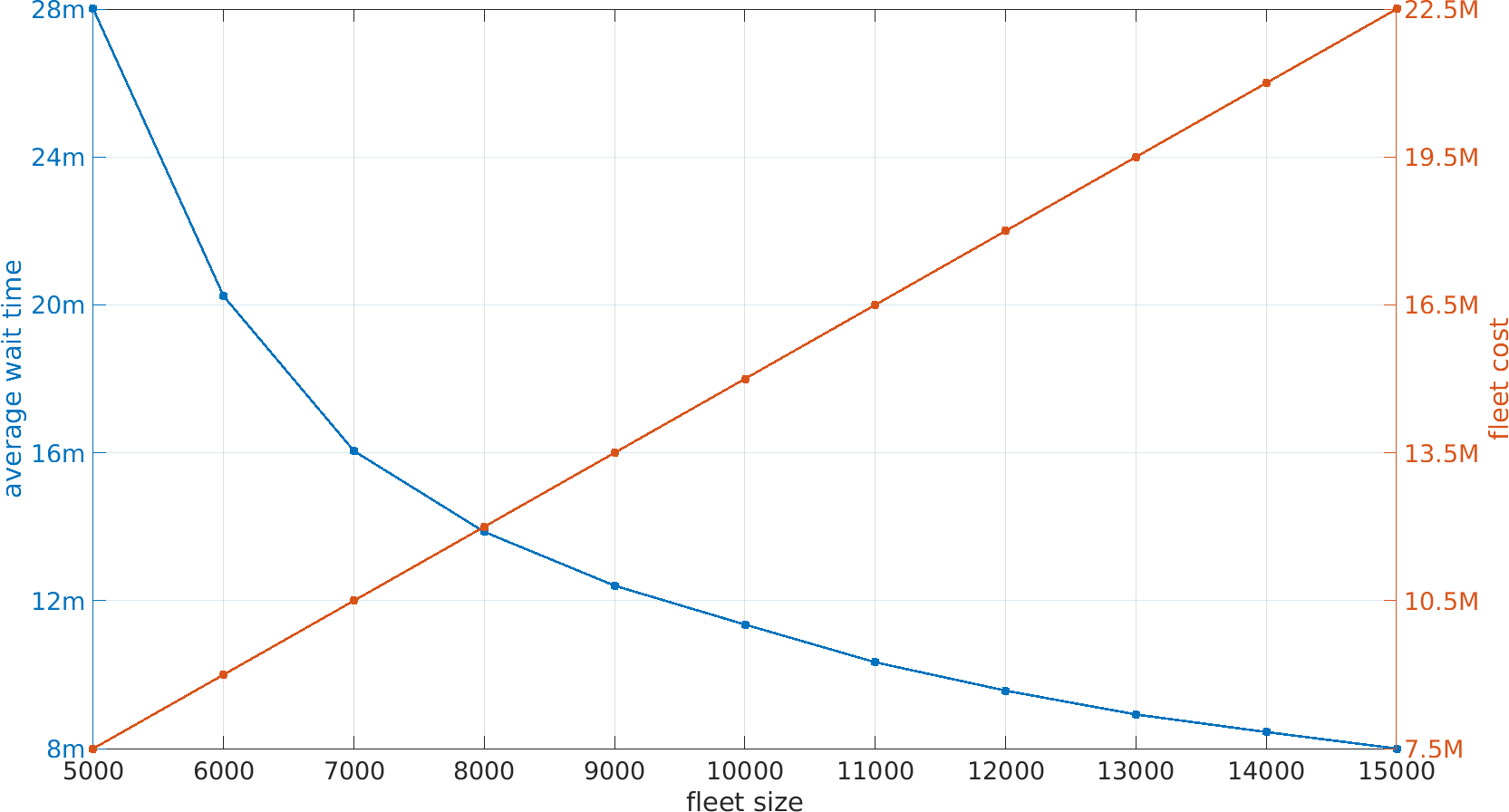} \put(20,45){$v_R = 1\,\mathrm{km}/\mathrm{h}$} \end{overpic} \,
\begin{overpic}[width=0.47\textwidth]{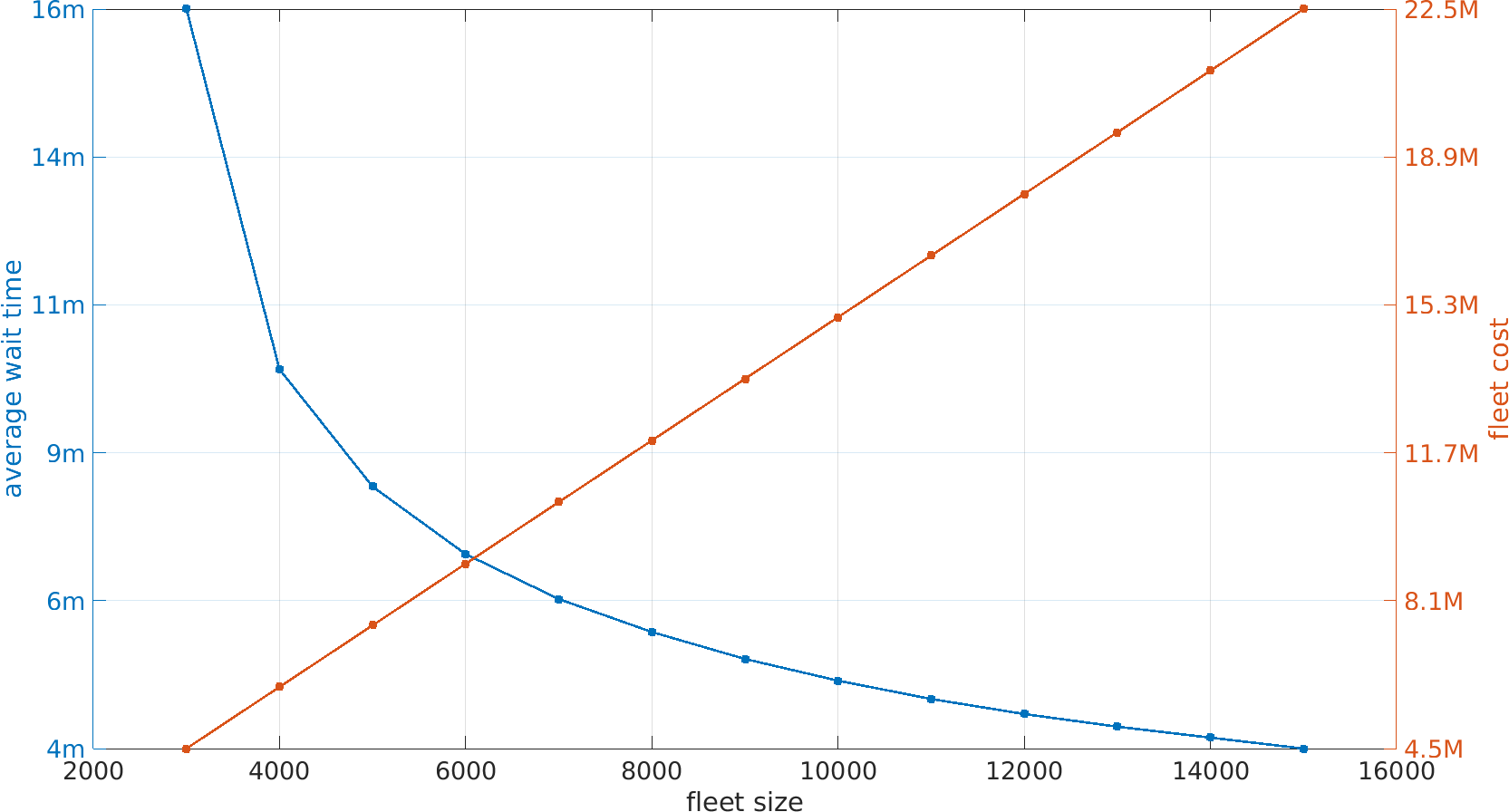} \put(20,45){$v_R = 2.5\,\mathrm{km}/\mathrm{h}$} \end{overpic} \\[3ex]
\begin{overpic}[width=0.47\textwidth]{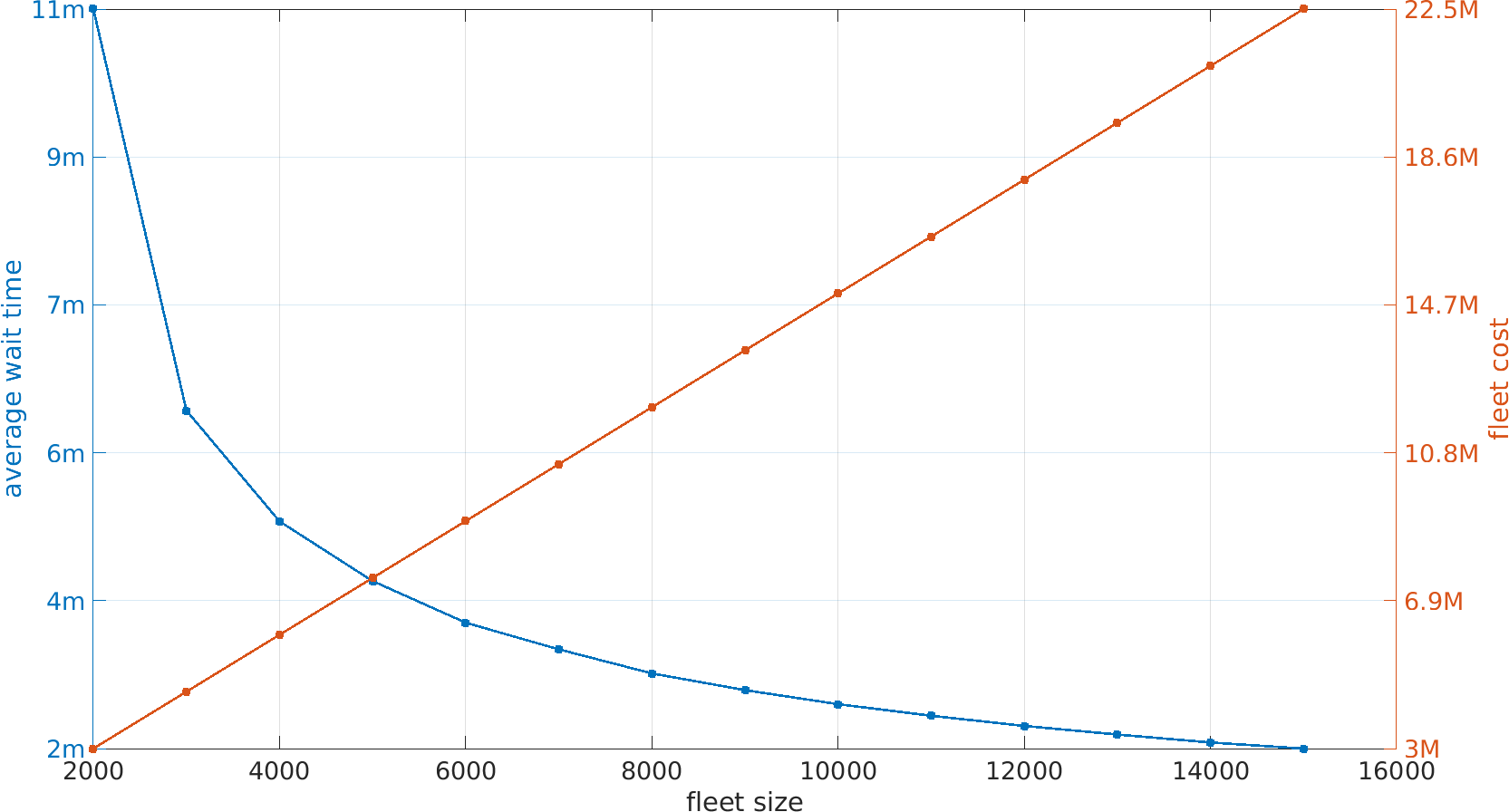} \put(20,45){$v_R = 5\,\mathrm{km}/\mathrm{h}$} \end{overpic} \,
\begin{overpic}[width=0.47\textwidth]{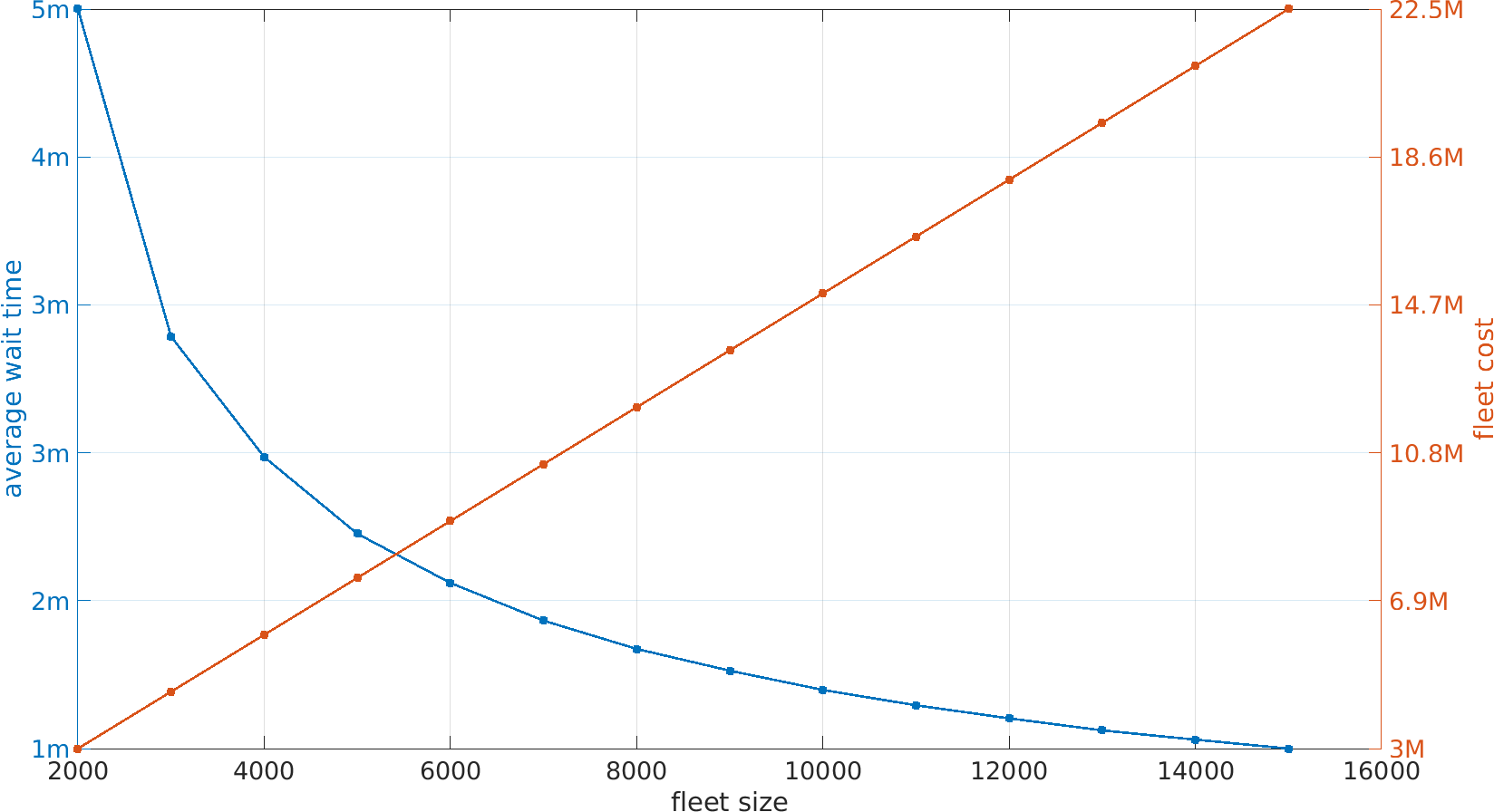} \put(20,45){$v_R = 10\,\mathrm{km}/\mathrm{h}$} \end{overpic}

	\caption{Comparison of user waiting times (blue, left $y$-axis) and fleet deployment cost (red, right $y$-axis, in SGD) in the online model for the four values of $v_R$ considered in our analysis. We note that the estimated cost of deploying non-autonomous vehicles corresponding to the average number of bikes in use (15,912 over one week) is approximately 9M SGD.}
	\label{fig:fleet_size_cost_wt}
\end{figure*}

\begin{table*}
\small
\begin{tabular}{l|r|r|r|r|r|r|r|r}
\multirow{2}{*}{day} &	\multirow{2}{*}{\makecell{number\\ of trips}} &	\multirow{2}{*}{\makecell{speed\\{} [km/h]}} &	\multirow{2}{*}{\makecell{number\\ of edges}} &	\multirow{2}{*}{\makecell{number\\ of matches}} &	\multicolumn{2}{c|}{unweighted matching runtime [s]} & \multicolumn{2}{c}{weighted matching runtime [s]} \\ 
 & & & & & network creation &	matching &	network creation &	matching \\ \hline
\multirow{4}{*}{Monday} &	\multirow{4}{*}{31058} &
		1 &	\numprint{74506594} &	\numprint{27527} &	1087 &	2.75 &	\numprint{1425} &	\numprint{274} \\
 &	 &	2.5 &	\numprint{209329664} &	\numprint{28597} &	\numprint{1084} &	9.26 &	\numprint{1480} &	\numprint{1401} \\
 &	 &	\numprint{5} &	\numprint{310495824} &	\numprint{29168} &	\numprint{1084} &	20.26 &	\numprint{1536} &	\numprint{2996} \\
 &	 &	\numprint{10} &	\numprint{381982617} &	\numprint{29557} &	\numprint{1097} &	30.67 &	\numprint{1570} &	\numprint{5567} \\ \hline
\multirow{4}{*}{Tuesday} &	\multirow{4}{*}{38030} &
		\numprint{1} &	\numprint{116851314} &	\numprint{34104} &	\numprint{1587} &	5.56 &	\numprint{1636} &	\numprint{539} \\
 &	 &	2.5 &	\numprint{334145396} &	\numprint{35324} &	\numprint{1251} &	16.67 &	\numprint{1811} &	\numprint{3066} \\
 &	 &	\numprint{5} &	\numprint{484232025} &	\numprint{35942} &	\numprint{1254} &	32.47 &	\numprint{1889} &	\numprint{6464} \\
 &	 &	\numprint{10} &	\numprint{583449630} &	\numprint{36382} &	\numprint{1280} &	53.81 &	\numprint{1799} &	\numprint{12552} \\ \hline
\multirow{4}{*}{Wednesday} &	\multirow{4}{*}{39645} &
		\numprint{1} &	\numprint{129124151} &	\numprint{35861} &	\numprint{1638} &	6.71 &	\numprint{1650} &	\numprint{666} \\
 &	 &	2.5 &	\numprint{369533687} &	\numprint{37061} &	\numprint{1294} &	23.27 &	\numprint{1771} &	\numprint{3679} \\
 &	 &	\numprint{5} &	\numprint{532800476} &	\numprint{37663} &	\numprint{1298} &	46.38 &	\numprint{1857} &	\numprint{9048} \\
 &	 &	\numprint{10} &	\numprint{639021770} &	\numprint{38078} &	\numprint{1318} &	65.81 &	\numprint{2018} &	\numprint{13566} \\ \hline
\multirow{4}{*}{Thursday} &	\multirow{4}{*}{39537} &
		\numprint{1} &	\numprint{133122014} &	\numprint{35659} &	\numprint{1614} &	6.54 &	\numprint{1664} &	\numprint{655} \\
 &	 &	2.5 &	\numprint{372796856} &	\numprint{36874} &	\numprint{1285} &	\numprint{23} &	\numprint{1761} &	\numprint{3438} \\
 &	 &	\numprint{5} &	\numprint{531145965} &	\numprint{37477} &	\numprint{1312} &	42.79 &	\numprint{1842} &	\numprint{7581} \\
 &	 &	\numprint{10} &	\numprint{634568937} &	\numprint{37867} &	\numprint{1313} &	63.26 &	\numprint{1873} &	\numprint{13612} \\ \hline
\multirow{4}{*}{Friday} &	\multirow{4}{*}{37143} &
		\numprint{1} &	\numprint{120938776} &	\numprint{33551} &	\numprint{1528} &	\numprint{6} &	\numprint{1737} &	\numprint{639} \\
 &	 &	2.5 &	\numprint{331723440} &	\numprint{34651} &	\numprint{1216} &	20.48 &	\numprint{1820} &	\numprint{3353} \\
 &	 &	\numprint{5} &	\numprint{468249862} &	\numprint{35242} &	\numprint{1253} &	38.5 &	\numprint{1844} &	\numprint{7226} \\
 &	 &	\numprint{10} &	\numprint{559596901} &	\numprint{35634} &	\numprint{1250} &	63.11 &	\numprint{1762} &	\numprint{13398} \\ \hline
\multirow{4}{*}{Saturday} &	\multirow{4}{*}{48747} &
		\numprint{1} &	\numprint{176832908} &	\numprint{44726} &	\numprint{1836} &	12.24 &	\numprint{2015} &	\numprint{1066} \\
 &	 &	2.5 &	\numprint{513206316} &	\numprint{45933} &	\numprint{1504} &	40.82 &	\numprint{2128} &	\numprint{6049} \\
 &	 &	\numprint{5} &	\numprint{784486744} &	\numprint{46529} &	\numprint{1513} &	86.78 &	\numprint{2348} &	\numprint{14628} \\
 &	 &	\numprint{10} &	\numprint{959889647} &	\numprint{46963} &	\numprint{1835} &	133.39 &	\numprint{2499} &	\numprint{24911} \\ \hline
\multirow{4}{*}{Sunday} &	\multirow{4}{*}{44666} &
		\numprint{1} &	\numprint{150748310} &	\numprint{40728} &	\numprint{1553} &	6.5 &	\numprint{1452} &	\numprint{760} \\
 &	 &	2.5 &	\numprint{441829507} &	\numprint{41834} &	\numprint{1361} &	28.36 &	\numprint{2126} &	\numprint{4933} \\
 &	 &	\numprint{5} &	\numprint{670507398} &	\numprint{42389} &	\numprint{1389} &	51.49 &	\numprint{2599} &	\numprint{9993} \\
 &	 &	\numprint{10} &	\numprint{814472181} &	\numprint{42789} &	\numprint{1897} &	69.28 &	\numprint{2630} &	17191
\end{tabular}
\caption{Runtimes for the maximum matching problem for finding optimal dispatching. Calculations were performed on a Lenovo P720 workstation with two Intel(R) Xeon(R) Silver 4116 CPUs and 256 GiB of RAM. All calculations shown here are single threaded. Number of edges refers to the total number of possible trip connections, that are represented as edges in a shareability network~\cite{Vazifeh2018}. Network creation runtime refers to the time needed to identify these edges and create an in-memory representation of them; this is achieved in this case by performing shortest path calculations (i.e.~a Dijkstra-search) on the OpenStreetMap path network; this step can be parellelised in a straightforward way and scales well on multiple cores. The matching runtime refers to the time needed to identify a maximum path cover on the shareability network, a computational problem that is best performed by solving a bipartite maximum matching on an augmented bipartite graph~\cite{Boesch1977}. The unweighted case is solved by the Hopcroft-Karp algorithm~\cite{HopcroftKarp}; code implementing it is available at \url{https://github.com/dkondor/graph\_simple}. The weighted case is solved using the Lemon library~\cite{lemon}, available at \url{https://lemon.cs.elte.hu/trac/lemon}. While the weighted case manifests extensive runtimes, these can be mitigated by limiting the size of the shareability networks by only including trip connections shorter than a given threshold, giving an approximate solution of the problem. Results in this case are further presented in Fig.~\ref{matching_runtimes_limit_fig} and Table~\ref{matching_runtimes_limit_tab}. All main results presented in the paper were calculated without such limits on connection times.}
\label{matching_runtimes}
\end{table*}

\begin{figure*}
	\centering
	\includegraphics{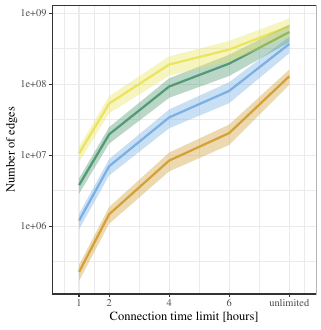}
	\includegraphics{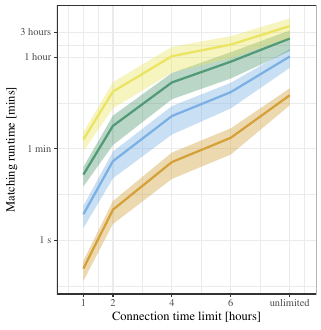}
	\includegraphics{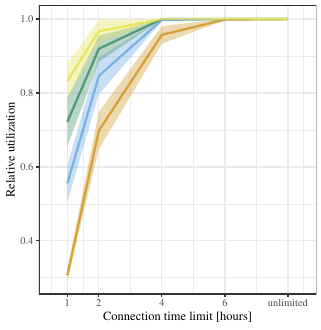}
	\includegraphics{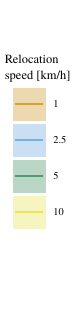}
	\caption{Comparison of approximate solutions to the minimum fleet problem. In these cases, maximum connection time (i.e.~time elapsed between the end of a trip and the start of the consecutive trip of the same vehicle) is limited to 1-6 hours. This results in significant reductions in the size of the shareability network (left panel) and the runtime to find the solutions (middle panel). At the same time, such limits can decrease the quality of the solutions, i.e.~resulting in larger estimated fleet sizes or lower estimated average utilization of vehicles. This effect is especially prominent for slow relocation speeds where some of the longer trips might be necessary to keep the fleet balanced during the day. Still, large gains in runtime can be realized even without degrading the solution quality by a large factor, as the results for a limit of 4 hours show, with the exception of $v_R = 1 \kph$. Figures here display average values for all days in the one week dataset used in this study. Note that the $y$-axis in the left and middle panels is logarithmic.}
	\label{matching_runtimes_limit_fig}
\end{figure*}

\begin{table*}
\small
\begin{tabular}{l|r|r|r|r|r|r|r|r|r|r}

\multirow{3}{*}{day} & \multirow{3}{*}{\makecell{number\\ of trips}} & \multirow{3}{*}{\makecell{speed\\{} [km/h]}} & \multirow{3}{*}{\makecell{Connection\\ time limit [h]}} & \multirow{3}{*}{\makecell{number\\ of edges}} & \multirow{3}{*}{\makecell{number\\ of matches}}  & \multicolumn{2}{c|}{Runtime [s]} & \multicolumn{3}{c}{Ratio to ideal solution} \\
& & & & & & \multirow{2}{*}{\makecell{network\\ creation}} & \multirow{2}{*}{matching} & \multirow{2}{*}{\makecell{number\\ of matches}} & \multirow{2}{*}{\makecell{fleet\\ size}} & \multirow{2}{*}{\makecell{average\\ utilization}} \\
& & & & & & & & & & \\ \hline
\multirow{4}{*}{Monday} & \multirow{4}{*}{31,058} & 1.0 & 4 & \numprint{5434016} & \numprint{27444} & \numprint{1450} & 13.3 & 0.997 & 1.024 & 0.977\\
 &  & 2.5 & \numprint{4} & \numprint{21938992} & \numprint{28597} & \numprint{1507} & \numprint{100} & \numprint{1} & \numprint{1} & 1\\
 &  & 5.0 & \numprint{2} & \numprint{12952501} & \numprint{29083} & \numprint{1568} & 67.8 & 0.997 & 1.045 & 0.957\\
 &  & 10.0 & \numprint{2} & \numprint{35005204} & \numprint{29557} & \numprint{1620} & \numprint{309} & \numprint{1} & \numprint{1} & 1\\ \hline
\multirow{4}{*}{Tuesday} & \multirow{4}{*}{38,030} & \numprint{1} & \numprint{4} & \numprint{7334512} & \numprint{34010} & \numprint{1721} & 24.9 & 0.997 & 1.024 & 0.977\\
 &  & 2.5 & \numprint{4} & \numprint{29642193} & \numprint{35301} & \numprint{1744} & \numprint{175} & 0.999 & 1.009 & 0.992\\
 &  & 5.0 & \numprint{2} & \numprint{17689197} & \numprint{35833} & \numprint{1731} & \numprint{113} & 0.997 & 1.052 & 0.95\\
 &  & 10.0 & \numprint{2} & \numprint{47430074} & \numprint{36353} & \numprint{1863} & \numprint{550} & 0.999 & 1.018 & 0.983\\ \hline
\multirow{4}{*}{Wednesday} & \multirow{4}{*}{39,645} & \numprint{1} & \numprint{4} & \numprint{8037980} & \numprint{35739} & \numprint{1814} & 30.6 & 0.997 & 1.032 & 0.969\\
 &  & 2.5 & \numprint{4} & \numprint{32474781} & \numprint{37061} & \numprint{1850} & \numprint{245} & \numprint{1} & \numprint{1} & 1\\
 &  & 5.0 & \numprint{4} & \numprint{87366729} & \numprint{37663} & \numprint{1986} & \numprint{1092} & \numprint{1} & \numprint{1} & 1\\
 &  & 10.0 & \numprint{2} & \numprint{51239910} & \numprint{38061} & \numprint{1952} & \numprint{745} & 0.9995 & 1.011 & 0.989\\ \hline
\multirow{4}{*}{Thursday} & \multirow{4}{*}{39,537} & \numprint{1} & \numprint{4} & \numprint{7823650} & \numprint{35490} & \numprint{1783} & 26.4 & 0.995 & 1.044 & 0.958\\
 &  & 2.5 & \numprint{4} & \numprint{32164439} & \numprint{36874} & \numprint{1830} & \numprint{217} & \numprint{1} & \numprint{1} & 1\\
 &  & 5.0 & \numprint{4} & \numprint{86571700} & \numprint{37477} & \numprint{1941} & \numprint{1038} & \numprint{1} & \numprint{1} & 1\\
 &  & 10.0 & \numprint{2} & \numprint{51931891} & \numprint{37789} & \numprint{1934} & \numprint{700} & 0.998 & 1.047 & 0.955\\ \hline
\multirow{4}{*}{Friday} & \multirow{4}{*}{37,143} & \numprint{1} & \numprint{6} & \numprint{16093247} & \numprint{33535} & \numprint{1744} & 70.2 & 0.9995 & 1.004 & 0.996\\
 &  & 2.5 & \numprint{4} & \numprint{27307712} & \numprint{34634} & \numprint{1729} & \numprint{178} & 0.9995 & 1.007 & 0.993\\
 &  & 5.0 & \numprint{4} & \numprint{74032545} & \numprint{35242} & \numprint{1853} & \numprint{857} & \numprint{1} & \numprint{1} & 1\\
 &  & 10.0 & \numprint{2} & \numprint{44317441} & \numprint{35562} & \numprint{1836} & \numprint{557} & 0.998 & 1.048 & 0.954\\ \hline
\multirow{4}{*}{Saturday} & \multirow{4}{*}{48,747} & \numprint{1} & \numprint{6} & \numprint{32011007} & \numprint{44723} & \numprint{2143} & \numprint{192} & 0.9999 & 1.001 & 0.999\\
 &  & 2.5 & \numprint{4} & \numprint{51958074} & \numprint{45933} & \numprint{2142} & \numprint{534} & \numprint{1} & \numprint{1} & 1\\
 &  & 5.0 & \numprint{4} & \numprint{143096994} & \numprint{46529} & \numprint{2274} & \numprint{2359} & \numprint{1} & \numprint{1} & 1\\
 &  & 10.0 & \numprint{4} & \numprint{291442345} & \numprint{46963} & \numprint{2390} & \numprint{7660} & \numprint{1} & \numprint{1} & 1\\ \hline
\multirow{4}{*}{Sunday} & \multirow{4}{*}{44,666} & \numprint{1} & \numprint{4} & \numprint{10391302} & \numprint{40621} & \numprint{1932} & \numprint{46} & 0.997 & 1.027 & 0.974\\
 &  & 2.5 & \numprint{4} & \numprint{42174871} & \numprint{41834} & \numprint{1970} & \numprint{361} & \numprint{1} & \numprint{1} & 1\\
 &  & 5.0 & \numprint{4} & \numprint{117642608} & \numprint{42389} & \numprint{2036} & \numprint{1487} & \numprint{1} & \numprint{1} & 1\\
 &  & 10.0 & \numprint{2} & \numprint{63230877} & \numprint{42747} & \numprint{2025} & \numprint{968} & 0.999 & 1.022 & 0.978\\
\end{tabular}
	\caption{Network sizes and runtimes for approximate solutions of the weighted minimum fleet problem. Cases displayed here were selected as the ones with minimal runtime where the ratio of average fleet utilization compared to the ideal solution (i.e.~the results in Table~\ref{matching_runtimes} is above $0.95$.}
	\label{matching_runtimes_limit_tab}
\end{table*}

\end{document}